\documentclass[journal]{IEEEtran}
\pdfoutput=1
\usepackage{color}
\usepackage{colortbl}
\usepackage{booktabs}
\usepackage{multirow}
\usepackage{tabularx}
\usepackage{amsthm}

\usepackage{mathrsfs}
\usepackage{bm}

\usepackage{threeparttable}

\usepackage{makecell}

\usepackage{flushend}

\usepackage{cite}
\ifCLASSINFOpdf
   \usepackage[pdftex]{graphicx}
   \usepackage{epstopdf}
   \usepackage{subfigure}
\else
   \usepackage[dvips]{graphicx}
\fi
\usepackage{amsmath}
\usepackage{amsfonts}
\usepackage{algorithm}  
\usepackage{algorithmicx}  
\usepackage{algpseudocode} 
\begin{document}
\title{Device-Bind Key-Storageless Hardware AI Model IP Protection: Joint PUF and Permute-Diffusion Encryption-Enabled Approach}

\author{Qianqian~Pan,~%~\IEEEmembership{Member,~IEEE,}
        Mianxiong~Dong,~\IEEEmembership{Member,~IEEE,}
        Kaoru~Ota,~\IEEEmembership{Member,~IEEE,}\\% <-this % stops a space
        ~and~Jun~Wu,~\IEEEmembership{Senior Membe,~IEEE}
\thanks{Qianqian Pan is with the School of Electronic Information and Electrical Engineering, Shanghai Jiao Tong University, Shanghai 200240, China and also with Shanghai Key Laboratory of Integrated Administration Technologies for Information Security, Shanghai 200240, China (E-mail: panqianqian@sjtu.edu.cn). }% <-this % stops a space
\thanks{Mianxiong Dong and Kaoru Ota are with Emerging Networks and Systems Laboratory, Department of Sciences and Informatics, Muroran Institute of Technology (E-mail: \{mx.dong, ota\}@csse.muroran-it.ac.jp).}% <-this % stops a space
\thanks{Jun Wu is with the Graduate School of Information, Production and Systems, Waseda University, Fukuoka, 808-0135 Japan (E-mail: junwu@aoni.waseda.jp), (Corresponding author: Jun Wu). }% <-this % stops a space

}


\markboth{\textgreater REPLACE THIS LINE WITH YOUR PAPER IDENTIFICATION NUMBER (DOUBLE-CLICK HERE TO EDIT)\textless}%
{Shell \MakeLowercase{\textit{et al.}}: Bare Demo of IEEEtran.cls for IEEE Journals}



\maketitle

\begin{abstract}
Machine learning as a service (MLaaS) framework provides intelligent services or well-trained artificial intelligence (AI) models for local devices. However, in the process of model transmission and deployment, there are security issues, i.e. AI model leakage due to the unreliable transmission environments and illegal abuse at local devices without permission. Although existing works study the intellectual property (IP) protection of AI models, they mainly focus on the watermark-based and encryption-based methods and have the following problems: (i) The watermark-based methods only provide passive verification afterward rather than active protection. (ii) Encryption-based methods are low efficiency in computation and low security in key storage. (iii) The existing methods are not device-bind without the ability to avoid illegal abuse of AI models. To deal with these problems, we propose a device-bind and key-storageless hardware AI model IP protection mechanism. First, a physical unclonable function (PUF) and permute-diffusion encryption-based AI model protection framework is proposed, including the PUF-based secret key generation and the geometric-value transformation-based weights encryption. Second, we design a PUF-based key generation protocol, where delay-based Anderson PUF is adopted to generate the derive-bind secret key. Besides, convolutional coding and convolutional interleaving technologies are combined to improve the stability of PUF-based key generation and reconstruction. Third, a permute and diffusion-based intelligent model weights encryption/decryption method is proposed to achieve effective IP protection, where chaos theory is utilized to convert the PUF-based secret key to encryption/decryption keys. Finally, experimental evaluation demonstrates the effectiveness of the proposed intelligent model IP protection mechanism.
\end{abstract}


% \fi
\IEEEpeerreviewmaketitle

\section{Introduction}
\label{section1}

\IEEEPARstart{M}{achine} Learning (ML) technologies play crucial roles in a wide range of application fields \cite{furuta2019fully, veres2019deep, sundaravadivel2017everything}, e.g. image processing, intelligent transportation, smart healthcare, etc. Due to constrained resources, it is difficult for local devices to train artificial intelligentce (AI) models alone for real-time data analysis and decision-making. Remote cloud and edge servers can provide local devices with well-trained models, namely machine learning as a service (MLaaS) \cite{wang2018facilitating, zhao2021veriml}. The well-trained intelligent models are considered as intellectual property (IP) of their owners, as the vast amount of resources consumed in the training process, including expensive hardware equipment, massive constructed data, and professional knowledge from experts \cite{xue2020intellectual}. For example, the GPT-3 intelligent model for human-like text generation of Google. 

The MLaaS faces the following security issues during model transmission and deployment: 1) Intelligent models leakage due to the unreliable transmission environment. 2) Illegal copy, redistribution, and abuse of intelligent models by malicious local devices without permission.  However, the security protection of intelligent models has not been well studied. Existing works are mainly based on traditional software security technologies \cite{yasaei2021gnn4ip, li2019prove}, including watermark-based and encryption-based algorithms, which have the following deficiencies. First, the watermark-based protection methods only provide passive verification afterward without the capacity to prevent the intelligent model IP actively. Second, encryption-based methods face low efficiency and insecure key storage/transmission issues. Besides, the development of hardware threat technologies (e.g. covert attacks and side-channel attacks) increases the risk of security-critical information exposure \cite{hu2020overview, pan2022side}, further denigrating the reliability of encryption-based methods. Third, the existing IP protection approaches are not device-bind, which is fatigued in preventing malicious users from copying and abusing AI models. Therefore, it is critically important to design the hardware security-based and device-bind scheme to achieve active AI IP protection.

To solve the above issues and challenges, physical unclonable function (PUF) and permute-diffusion encryption technologies are introduced. PUF is an important hardware-intrinsic security primitive, which uses random manufacturing variations to generate a unique unforgeable device fingerprint \cite{herder2014physical}. With features of permanent, reliability, and unpredictability, PUF has the capability to realize device-specific intelligent model IP protection. Permute-diffusion encryption technology is widely used in image encryption, with characteristics of computational efficiency and suitable for large-scale data \cite{singh2022towards}. To overcome the conflict between the intensive computing of traditional cryptography-based authentication algorithms and the limited resources of local devices, permute-diffusion encryption technology is introduced to realize fast encryption/decryption in the protection of AI model IP.

To realize active, device-bind, and efficient AI model IP protection, we propose a PUF and permute-diffusion encryption technologies empowered approach. The contributions of this work are summarized as follows:
\begin{itemize}
 \item We propose a PUF and permute-diffusion encryption technologies empowered AI IP protection framework, including the PUF-based secret key generation scheme and the geometric-value transformation-enabled weights encryption scheme. This framework supports the pay-per-device MLaaS and provides tamper-proof security protection against physical attacks. 
 %We analyze the application scenario and the threat models, and design the intelligent model protection method based on PUF-driven encryption and   
 
 \item A PUF-based key generation protocol is designed, where delay-based Anderson PUF is adopted to derive the device-bind secret keys. In the designed protocol, convolutional coding and convolutional interleaving technologies are combined to improve the stability of key generation and reconstruction in AI model protection.% 
 
 \item To achieve the fast and efficient encryption and decryption of AI models, permute and diffusion-based weights encryption/decryption mechanism is proposed. In this mechanism, chaos theory is adopted to convert the PUF-based secret key to encryption/decryption keys. Positions and values of model weights are transformed to protect AI model IP.

\end{itemize}

The remainder of this paper is organized as follows. Related work is discussed in Section \ref{sec:related}. The PUF and permute-diffusion encryption technologies empowered the AI IP protection framework is proposed in Section \ref{sec:framework}. In section \ref{sec:puf}, a PUF-based key generation scheme for AI IP protection is implemented. Section \ref{sec:image} presents the PUF and permute-diffusion encryption technologies-based AI weights encryption/decryption method. Section \ref{sec:eva} discusses the experiments. Finally, Section \ref{sec:concl} concludes this paper.

\section{Related Work}
\label{sec:related}

\subsection{Intelligent Model IP Protection}
The protection of AI model IP has attracted attention recently since its high cost and crucial value. Watermark-based and encryption-based algorithms are widely utilized in intelligent model IP protection.

Watermark-based IP protection schemes embed a watermark into the intelligent models to claim ownership. Uchida \textit{et al.} in \cite{uchida2017embedding} propose a digital watermark method-based AI IP protection scheme, where watermarks are embedded into parameters of AI in the model training process. The authors in \cite{zhang2018protecting} study three watermark generation methods and design a watermark implanting mechanism to protect the IP of DNN models. The proposed IP protection mechanism is capable to verify the model ownership remotely. Adi \textit{et al.} \cite{adi2018turning} investigate a watermark-based IP protection method for DNN in a black-box manner, where a backdoor is utilized as a watermark key to protect the IP of intelligent models for general classification tasks. Although these watermark-based algorithms protect the IP of intelligent models, they change the weights of models and reduce their performance. Besides, the watermark-based algorithms provide passive protection of intelligent models and only verify model ownership afterward rather than proactive protection.

Encryption-based AI IP protection schemes adopt encryption algorithms to protect intelligent model weights. The authors in \cite{cai2019enabling} design an effective neural network weight encryption framework, which includes a sparse fast gradient encryption scheme and a runtime encryption scheduling sheme to protect model IP. Chakraborty \textit{et al.} design a hardware-assisted IP protection of deep learning models, where DNN is trained as a function of the secret key and only ledge users with the embedded secret key in hardware can access the correct intelligent model \cite{chakraborty2020hardware}. However, these existing encryption-based AI IP protection schemes need to transmit secret keys in unreliable environments and store them in local devices, which are vulnerable to side-channel and reverse attacks. In addition, the existing AI IP protection schemes are not device-bind, which cannot solve the abuse of intelligent models after they are deployed at local devices.

\subsection{Physical Unclonable Function}
As a security primitive, the security of PUF stems from the physical microstructure of the hardware, making it resistant to multiple intrusive attacks. PUF uses simple and low-power circuits to realize security protection, which has been applied in multiple areas, e.g. key generation, IP protection, and authentication. Usmani \textit{et al.} in \cite{usmani2018efficient} investigate the efficient PUF-empowered key generation in FPGA, where the authors improve the design of the delay-based weak PUF and propose the per-device configuration. Evaluation of the FPGA hardware platform demonstrates the effectiveness of the designed PUF-based key generation methodology. The authors in \cite{zhang2015puf} propose a binding scheme for the IP protection of FPGA based on PUF and finite-state machines technologies. In this scheme, a pay-per-device licensing is designed which supports IP purchasing based on users' demands. Gope \textit{et al.} in \cite{gope2021comparative} investigate the PUF-based security protocols in the Internet of Things (IoT), which introduce and analyze multiple PUF-based authentication protocols for IoT in an ideal and noisy environment.

Seldom works apply PUF technology to protect intelligent models protection. The authors in \cite{guo2018puf} design a PUF-enabled pay-per-device IP protection scheme for convolutional neural network (CNN) models, which obfuscates the trained CNN models with the assistance of PUFs. However, this scheme only focuses on the CNN model instead of the general intelligent model protection. Moreover, this scheme sometimes needs to adjust CNN model weights to obtain stable challenge-response pairs of the PUF, which has affection on the model accuracy.

\subsection{Image Encryption}
Image encryption technologies aim to achieve efficient data protection for large-scale images based on space, transformation, and compressed sensing approaches. The authors in \cite{guan2019chaotic} propose a chaotic theory-based image encryption algorithm, where a hyper chaos system is designed to generate a random sequence for image encryption. Besides, this algorithm transforms the image into the frequency domain and the deoxyribonucleic acid encoding technology is used for data protection. Mario \textit{et al.} in \cite{preishuber2018depreciating} design a block scrambling-based image encryption method. In this method, images are divided into multiple blocks. Then, geometric and color transformations of the images are applied to protect image information. The authors in \cite{liu2022image} focus on the security and efficiency issues of image transmission. They propose a compressive sensing and hyperchaotic system empowered algorithm, which provides security protection while reducing data size. Tatsuya \textit{et al.} investigate a block scrambling-driven image encryption scheme in \cite{chuman2018encryption}, where an encryption-then-compression framework is designed for JPEG images. 

Some works investigate image encryption-based AI model protection. For example, Lin \textit{et al.} propose a chaotic map theory-based weights encryption method for deep neural networks (DNN) models in \cite{lin2020chaotic}, where model weights are chastized by exchanging positions. Although the image encryption-based IP protection methods are efficient, they are not combined with the PUF technology to achieve device-bind protection.

\section{PUF and Permute-Diffusion Encryption Empowered AI IP Protection Framework}
\label{sec:framework}

\subsection{Scenario and Threat Model of AI IP}
\begin{figure}[t]
\centering
\includegraphics[width=3.3in]{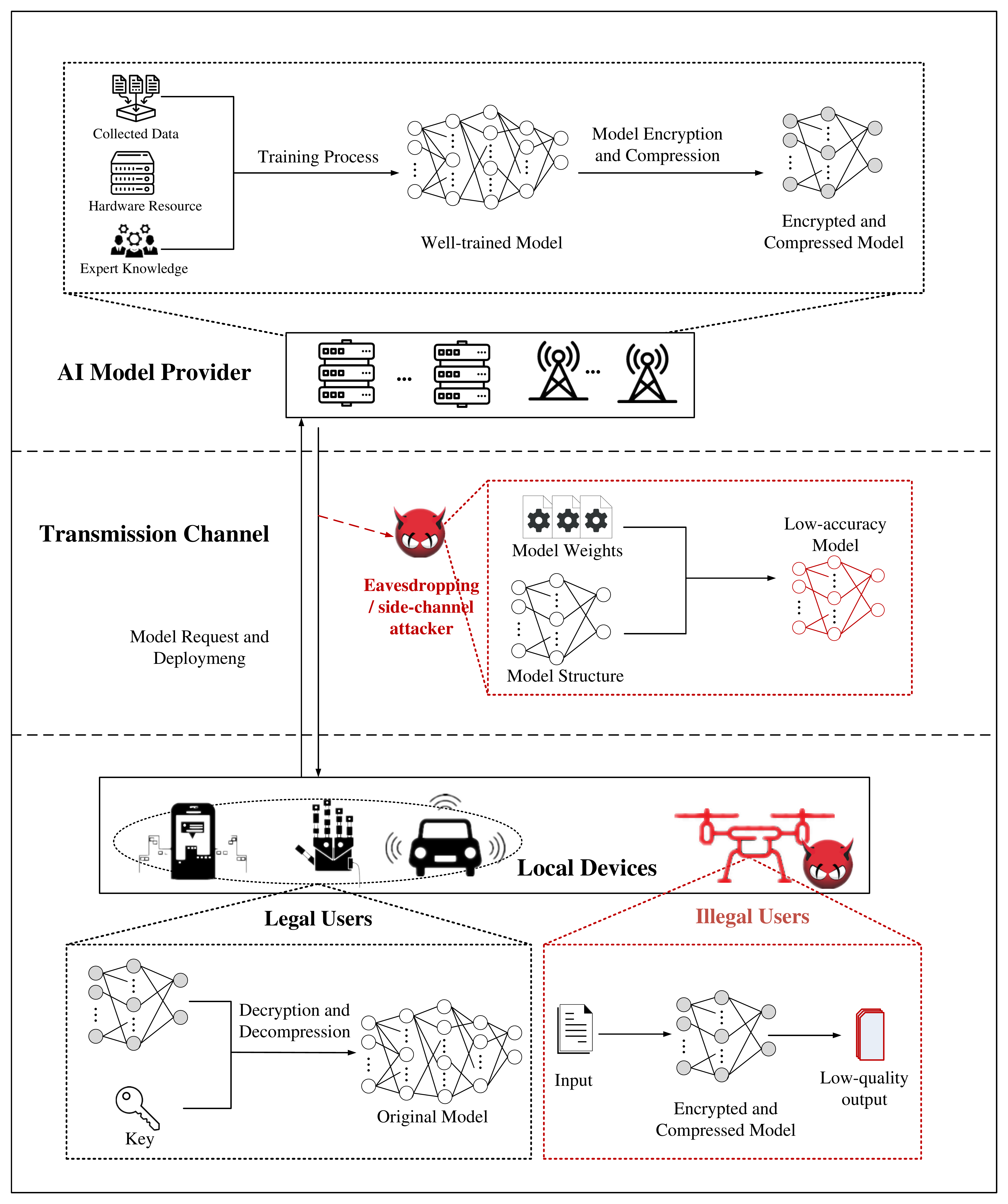}
\centering
\caption{Scenario for threat model of AI IP in the MLaaS.}
\label{Scenario}
\end{figure}

The scenario and threat models of AI IP are analyzed. As shown in Fig. \ref{Scenario}, the MLaaS architecture mainly involves three aspects, i.e. intelligent model provider, transmission channel, and local devices.

The intelligent model providers utilize collected data, hardware resources, and expert knowledge to develop high-precision intelligent models for multiple novel applications. To preserve user privacy, well-trained intelligent models are usually deployed on local devices rather than the remote cloud server. There are two types of threats in the process of model deployment: 
\begin{itemize}
\item{Model eavesdropping during transmission:} Due to the untrusted communication environment, intelligent models are exposed to adversaries during transmission. Adversaries monitor the model transmission channel and eavesdrop on the well-trained model information. 

\item{Model extraction and abuse at local devices:} After being deployed on local devices, intelligent models are vulnerable to illegal users. With the development of reverse engineering and side-channel attacks, illegal users may extract model weights and structures to reconstruct the AI model. More seriously, malicious users may abuse the well-trained intelligent models, i.e. copying the model and forwarding it to other colluded devices. 
\end{itemize}

To protect the IP of intelligent models, it is necessary to design an intelligent model encryption method to avoid direct exposure of model information to adversaries. Even if eavesdroppers get the model data packet, they cannot leverage the model to achieve high-precision performance because they do not have the corresponding decryption key. Moreover, the intelligent model encryption method should be device-bind to solve the model-abusing issues at local devices. Namely, the private key of one local device cannot be used to decrypt encrypted models on the other devices. 

\subsection{AI IP Protection Framework}
\begin{figure}[t]
\centering
\includegraphics[width=3.5in]{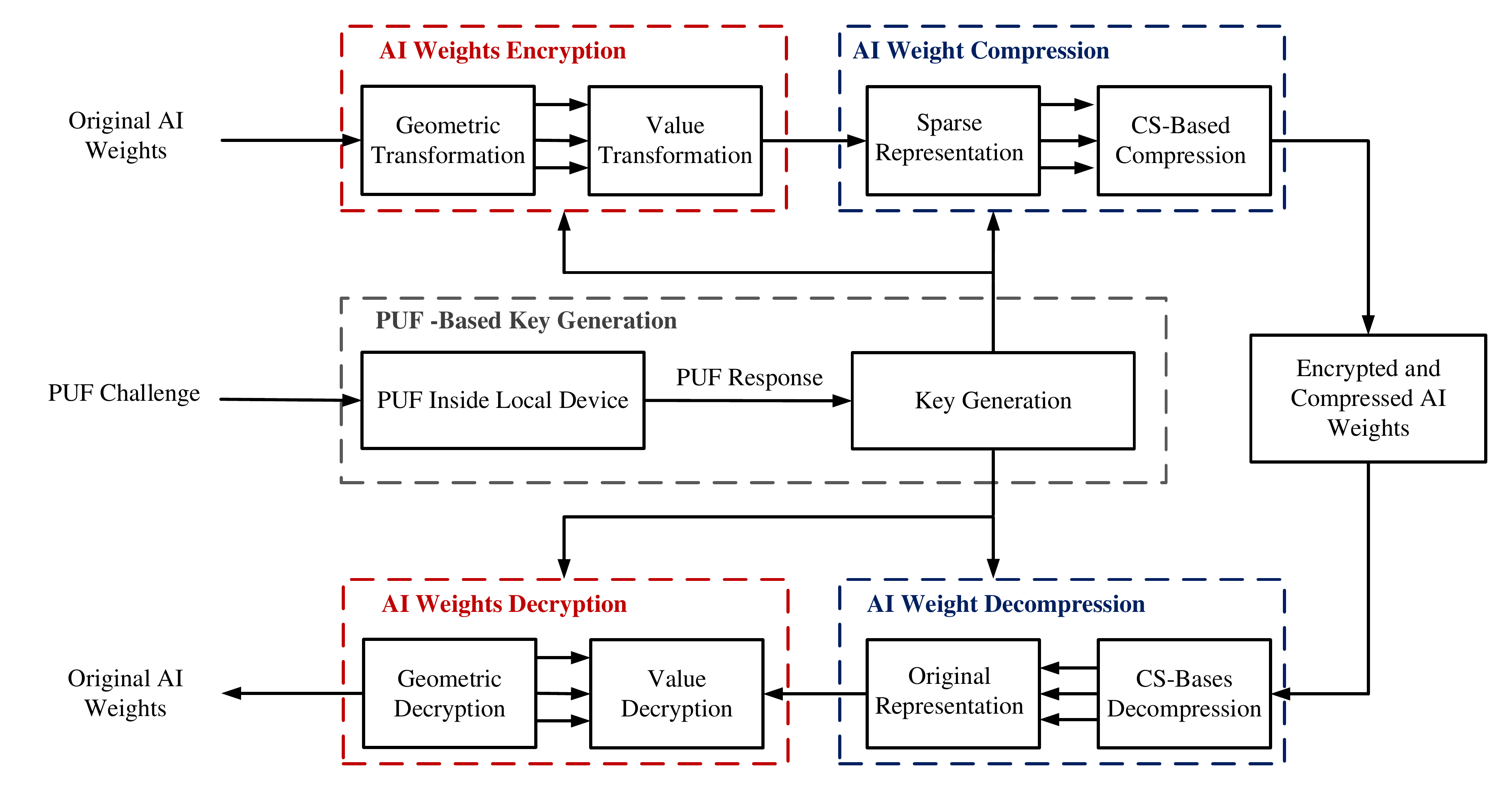}
\centering
\caption{Design principle of the PUF and permute-diffusion encryption-enabled AI IP protection.}
\label{framework}
\end{figure}

To realize the secure transmission and the device-bind deployment of intelligent models, the intelligent model IP protection framework is established based on PUF and permute-diffusion encryption technologies. As shown in Fig. \ref{framework}, the proposed AI model IP protection framework consists of the following three parts, i.e. PUF-based key generation, permute-diffusion model encryption, and intelligent model decryption.

With features of reliability, unpredictability, and unclonability, PUF is utilized to generate secret keys in AI model IP protection. According to the PUF challenges, PUF inside local devices outputs device-specific responses due to the manufacturing process. Then, the PUF response is applied to generate secret keys for intelligent model encryption and decryption. The generated secret keys only rely on the inherent physical variation and need not be stored in the nonvolatile memory. The proposed PUF-based AI IP protection supports on-device key generation, which is resistant to side-channel attacks and reverse attacks.

According to the secret key generated based on the PUF response, intelligent model providers encrypt model weights to protect their intellectual property. Permute-diffusion encryption algorithms are utilized to realize effective intelligent model protection, which includes geometric and value transformation operations. Then, encrypted intelligent models are transferred via untrusted channels to local devices that subscribe to the intelligent services. Along with the encrypted model package, the PUF challenge is also transmitted to local users for on-device secret key generation.

After receiving the encrypted model and challenge information, the local device adopts the PUF module inside the devices to obtain the corresponding response toward the challenge. Then, secret keys are generated based on the obtained PUF response. Since the challenge and response mapping of PUF is unique, only the specific legal local device is possible to generate the correct response and secret key. Based on the generated secret key, the encrypted intelligent model is decrypted, which consists of the value and geometric decryption operations.  

\section{PUF-Based Key Generation for AI Model IP Protection}
\label{sec:puf}
To realize the device-bind AI model IP protection, a PUF-based key generation scheme is proposed in this section. 

\subsection{Anderson PUF-Based Key Generation}

PUF is formulated as a mapping function from finite challenge space $\mathcal{C}$ to the response space $\mathcal{R}$. The response $R_i$ for the challenge $C_i$ is random and unpredictable, but the response is unchanged for the same challenge. It is impossible for an adversary to reconstruct a PUF that satisfies all the challenge-response mappings since the uncontrollability of process variation. These properties make PUF feasible and suitable for key generation, where challenge-response pairs (CRP) are utilized to generate secret keys on devices.

\begin{figure*}[t]
\centering
\includegraphics[width=7in]{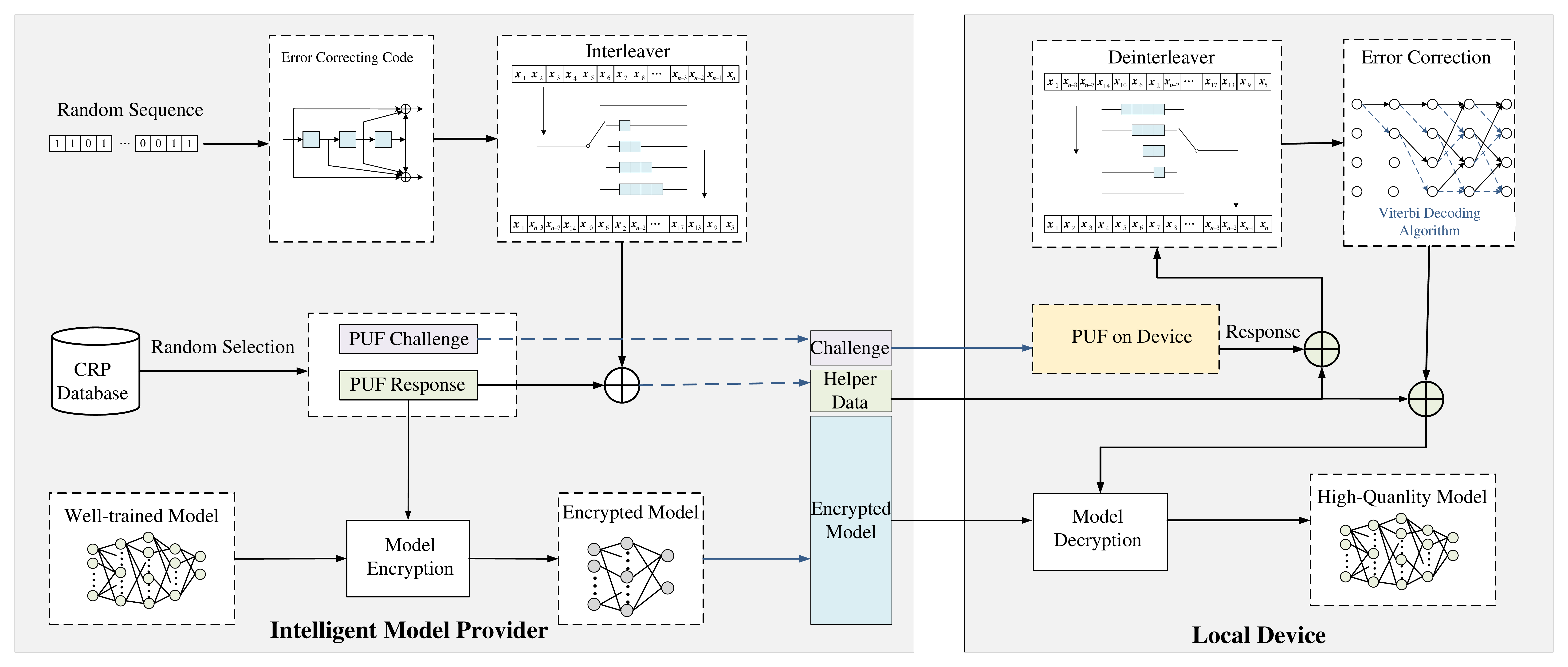}
\centering
\caption{PUF-Based key generation for intelligent model IP protection.}
\label{PUF}
\end{figure*}

The proposed PUF-based key generation approach is shown in Fig. \ref{PUF}. The PUF inside local devices of this paper adopts the delay-based PUF. As a classic PUF structure, Anderson PUF is utilized for key generation \cite{usmani2018efficient, zhang2015puf}. As shown in Fig. \ref{PUF}, the Anderson PUF consists of two shift registers, one flip flop, and multiple multiplexers. The inputs of these two shift registers are $0{\text{x}}5555$ and $0{\text{x}}AAAA$, respectively, whose outputs are respectively connected to different multiplexers as selectors. Between these multiplexers, there are multiple multiplexers as a delay chain. Although the outputs of both shift registers cannot be “1” at the same time, there is a time duration when both the input and selector of the top multiplexer are “1”  due to the delay chain. Thus, there exists a pulse in the output of the multiplexer, which is connected to the flip flop. If the duration and amplitude of the pulse exceed the preset value of the preset input, the output will be “1”, otherwise "0". The output of the Andrson PUF is unique for each device due to the manufactured difference, which can be applied as the device-bind secret key. 

The PUF modules on devices are sensitive to environmental variation (e.g. temperature and voltage), which may result in different responses of the PUF for the identical challenge. To improve the stability of the PUF module inside devices, a convolutional coding and convolutional interleaving empowered fuzzy extractor is designed. Convolutional code is one type of error-correcting code technology in telecommunication, which generates supervisory symbols by sliding boolean polynomial functions along information symbols. The convolutional coder utilized in the PUF-based key generation approach is expressed as $(N,L,M)$, where $N$, $M$, and $L$ are represented as the number of input bits, shift register stages, and output bits respectively. For example, the convolutional coder in Fig. \ref{PUF} can be presented as $(1,3,3)$. The designed convolutional coder is represented as a generator sequence $(\bm{g}^{(1)} ,\bm{g}^{(2)} ,\cdots, \bm{g}^{(L)} )$, where $\bm{g}^{(l)}$ is the impulse response for the impulse input signal $\delta=(1,0,0,0,\cdots)$ and
\begin{equation}
  \bm{g}^{(l)} = (g_0^{(l)}, g_1^{(l)},g_2^{(l)},\cdots, g_M^{(l)}), \forall l \in \{1,2,\cdots, L\}.
  \label{g}
\end{equation}
In (\ref{g}), $g_m^{(l)} \in  \{0,1\}$ for $m=0,1,2,\cdots,M$ and $l=1,2,\cdots, L$.  

Suppose the local devices set of the proposed AI IP protection system is denoted as $\mathcal{D}$. For the local device $i \in \mathcal{D}$, the response of PUF (i.e. secret key) is $\bm{k}_{i}=(k_{i,1}, k_{i,2} \cdots k_{i,B})$, where $B$ is the number of key bits. After the convolutional coding, the outputs are 
\begin{equation}
   \bm{v}_{i}^{(l)} = \bm{k}_{i} \otimes \bm{g}^{(l)}, \quad \forall l \in \{1,2,\cdots, L\},
\end{equation}
where $\otimes$ is the discrete convolution operation. $\bm{v}_{i}^{(l)} \in \{0,1\}^{T}$ where $T=B+M-1$. For each component in $\bm{v}_{i}^{(l)}$, 
\begin{equation}
  \begin{split}
   {v}_{i,j}^{(l)} &= k_{i,j} g_0^{(l)} + k_{i,j-1} g_1^{(l)} + \cdots + k_{i,j-M} g_M^{(l)} \\
   \forall j&= 1,2, \cdots,T,
  \end{split}
\end{equation}
where $k_{i,j}=0$ for $j<1$ and $j>B$. Then, the coded secret key for intelligent model IP protection is the combination of all the outputs, expressed as 
\begin{equation}
   \bm{v}_{i} = ( \{{v}_{i,t}^{(0)} {v}_{i,t}^{(1)} \cdots {v}_{i,t}^{(L)}\})_{t=1}^{T}.
\end{equation}
The commonly applied convolutional decoder is based on the Viterbi decoding algorithm. The error-correcting capability of the convolutional coding algorithm is related to the minimum free distance, defined as 
\begin{equation}
  \begin{split}
   d_{\text{free}} &= \min_{\bm{k}_{i}, \bm{k}'_{i}}\{ d(\bm{v}_{i},\bm{v}'_{i})|\bm{k}_{i} \neq \bm{k}'_{i}\}\\
   & \overset{\text{(a)}}{=} \min_{\bm{k}_{i}} \{ hw(\bm{v}_{i}) |  \bm{k}_{i} \neq 0\}
  \end{split}
\end{equation}
where $hw(\cdot)$ is the hamming weight function and (a) is because the convolutional code is a linear code. Therefore, the error-correcting capability of convolutional code $(N,L,M)$ is measured by
\begin{equation}
   r = \lfloor \frac{d_{\text{free}}-1}{2} \rfloor = \lfloor \frac{\min_{\bm{k}_{i}} \{ hw(\bm{v}_{i}) |  \bm{k}_{i} \neq 0\}-1}{2} \rfloor,
\end{equation}
where $r$ is the number of error bits the convolutional code $(N,L,M)$ can correct.

Since the variation of temperature and voltage, the PUF outputs are vulnerable to burst interference. To mitigate this issue, a convolutional interleaver is introduced, which has the ability to disperse burst errors into independent single errors. Then, error-correcting code technology is utilized to correct error bits. The interleaver consists of $S$ shift registers, whose caching sizes are $0, Q, \cdots, (S-1)Q$ respectively. Since the various size of shift registers, different delays of symbols are caused. Shift registers $1$ to $S$ receive the symbol from the convolutional coder sequentially and output their cached symbols iteratively. The corresponding deinterleaver has a symmetrical structure with the interleaver. The caching size of $S$ shift registers in deinterleaver are $(S-1)Q,  (S-2)Q, \cdots, Q, 0$ respectively. The output of the convolutional interleaver is considered as helper data, which is utilized to assist to obtain stable PUF-based secret keys in a noisy environment.

\subsection{PUF-Based Key Generation Protocol}

The PUF-based key generation protocol includes two phases, i.e. registration phase and intelligent model deployment phase, which are described as follows.

\subsubsection{Registration Phase}
The registration phase is carried out on secure channels. In this phase, there are the following five steps:
\begin{enumerate}
  \item Local device $i \in \mathcal{D}$ sends its ID $I_i$ to the intelligent models provider.
  \item Intelligent models provider generates a sequence of challenges $\bm{C}_i=\{C_{i,1}, C_{i,2}, \cdots, C_{i, z}\}$ and this challenge sequence to the local device.
  \item The local device generates response sequence $\bm{R}_i=\{R_{i,1}, R_{i,2}, \cdots, R_{i, z}\}$, where $R_{i,j}=PUF_i(C_{i,j})$ for $j=1,2,\cdots, z$. The local device sends CRP sequence $(\bm{C}_i,\bm{R}_i)=(\{C_{i,1}, R_{i,1}\}, \{C_{i,2}, R_{i,2}\}, \cdots, \{C_{i,z}, R_{i,z}\})$ to the model provider. 
  \item The intelligent model provider store the received CRPs $(\bm{C}_i,\bm{R}_i)$ in its database. Besides, the provider calculates helper data $\bm{hd}_i=\{hd_{i,1}, hd_{i,2},\cdots, hd_{i,z}\}$ based on the revieced responses $\bm{R}_i$ and the designed fuzzy extractor. Then, challenge and helper data pairs $(\bm{C}_i,\bm{hd}_i)=(\{C_{i,1}, hd_{i,1}\}, \{C_{i,2}, hd_{i,2}\}, \cdots, \{C_{i,z}, hd_{i,z}\})$ are sent to the local device.
  \item The local device store $(\bm{C}_i,\bm{hd}_i)$ in its memory.
\end{enumerate}

\subsubsection{Intelligent Model Deployment Phase}
The designed intelligent model deployment phase provides security protection for AI model IP. This phase consists of five steps, which are detailed as follows:
\begin{figure*}[t]
\centering
\includegraphics[width=7in]{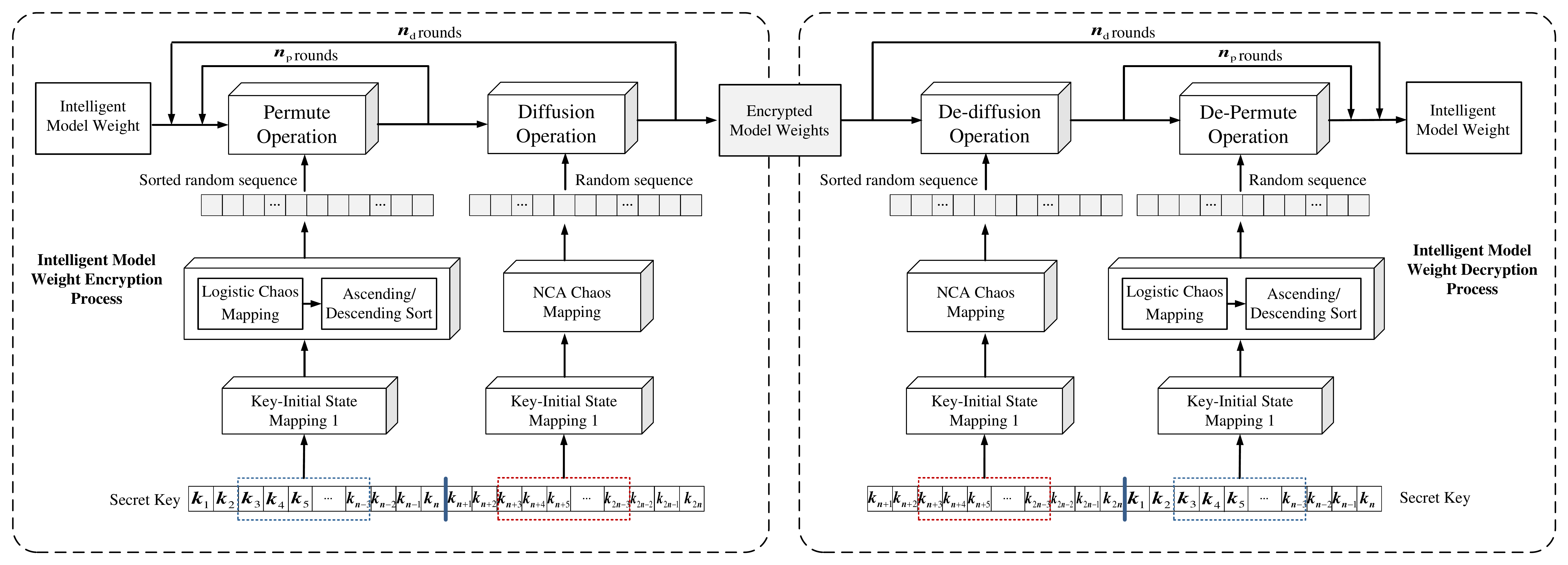}
\centering
\caption{Permute-diffusion encryption empowered intelligent model IP protection.}
\label{enc}
\end{figure*}
\begin{enumerate}
  \item The local device requests for the intelligent model from the provider by sending message $\mathcal{M}_{\text{d}1}: \{I_i, N_{\text{d}}\}$, where $N_{\text{d}}$ is the random number.
  \item Based on the ID $I_i$ of the local device, the intelligent model provider checks whether the device subscribes to the intelligent model service or not. If the device $I_i$ has subscribed to the service from the provider and paid enough fees, then a CRP $\{C_{i}, R_{i}\}$ is selected from the database of providers randomly. For mutual authentication with the local device, the provider sends the following to the local device:
\begin{equation}
   \mathcal{M}_{\text{p}1}: \{C_i,  R_i \otimes N_{\text{p}}, hash(N_{\text{d}} || R_i) \},
\end{equation}
where $N_{\text{p}}$ the random number selected by provider and $hash(\cdot)$ is the hash function.
  \item Based on the $\mathcal{M}_{\text{p}1}$, the local device calculates PUF response $R'_i=PUF_i(C_i)$ and gets helper data $hd_i$ from its memory. Then, convolutional decoder and deinterleaver are utilized to correct $R'_i$ with the assistance of helper data $hd_i$ and obtain the corrected PUF response $R_i$. Next, the local device check the $hash(N_{\text{d}} || R_i)$. If the $hash(N_{\text{d}} || R_i)$ is verified, it means the intelligent model provider is authenticated by the local device. After that, the local device transmits the message $\mathcal{M}_{\text{d}2}: \{hash(N_{\text{p}} || R_i)\}$ to the provider.
  \item The model provider verifies the message $hash(N_{\text{p}} || R_i)$ to authenticate the identity of the local device. If the device is authenticated, the provider encrypts the intelligent model with the secret key $k_i=R_i$ and sends $\mathcal{M}_{\text{p}2}: ENC( \bm{w}_{i}, k_i)$ to the local device, where $ENC_{k_i}(\cdot)$ is the encryption function for intelligent models and $\bm{w}_{i}$ denotes the intelligent model.
  \item After receiving the encrypted intelligent model, the local device decrypts it based on the corrected PUF response $R_i$, i.e. $\bm{w}_{i}=DEC(ENC( \bm{w}_{i}, k_i), R_i)$.
\end{enumerate}

In the intelligent model deployment phase, the local device sends requests in step 1. In steps 2 to 4, the model provider and the local device authenticate each other. Since the size of the encrypted intelligent models is large, it is resource-consuming for model transmission. Therefore, the provider verifies the devices' identity first and only transmits encrypted intelligent models to legal devices, which relieves the pressure of data transmission and filters malicious requests from illegal devices. Besides, the local device authenticates the model provider to ensure the reliability of received messages and intelligent models.

\section{PUF and Permute-Diffusion Encryption Empowered Intelligent Model IP Protection}
\label{sec:image}
The intelligent model weights encryption and decryption are proposed in this section based on the PUF and permute-diffusion encryption technologies.

\subsection{Permute-Diffusion based AI Model Encryption/Decryption Mechanism}
The workflow of the proposed mechanism is shown in Fig. \ref{enc}, which consists of permuting and diffusion two types of operations. The permute operation and the diffusion operation are responsible for the geometric transformation and the value transformation of the weight elements, respectively. Compared with number theory-based encryption approaches (e.g. AES and DES), permute-diffusion based image encryption method is more computationally efficient. 

Suppose an AI model $\mathcal{A}: \mathcal{X}^n \rightarrow \mathcal{Y}^m$ and its connection weights can be denoted as $\mathbf{w}=(\bm{w}^{(1)}, \bm{w}^{(2)},\cdots, \bm{w}^{(N_{\text{w}})})$, where $N_{\text{w}}$ is the number of layers of the intelligent model. For the layer $j \in \{1,2, \cdots, N_{\text{w}}\}$, $\bm{w}^{(j)} \in \mathcal{R}^{L_{j} \times L_{j-1}}$, where $L_j$ is the output size of the layer $j$ and $L_0$ is the input size of the whole intelligent model. The intelligent model encryption process is executed on the model provider and includes the following four steps for encryption of each layer weight $\bm{w}^{(j)}$ for $j=1,2, \cdots, N_{\text{w}}$:
\begin{enumerate}
  \item Convert the multi-dimensional weight matrix $\bm{w}^{(j)}$ to one dimension as a sequence $\bm{w}'^{(j)}$. 
  \item Perform permute operations $n_{\text{p}}$ rounds on the weight sequence $\bm{w}'^{(j)}$ to exchange elements positions of the weight $\bm{w}'^{(j)}$. These permute operations are executed based on the secret keys and details are provided in the next subsection.
  \item Diffusion operation is executed to modify values of elements in the weight $\bm{w}'^{(j)}$. Then, go back to step 2 to perform the $n_{\text{p}}$-round permute operations and $1$-round diffusion operation repeatedly for $n_{\text{d}}$ times.
  \item Reconvert the one-dimensional $\bm{w}'^{(j)}$ to multiple dimensions as the encrypted weight, denoted as $\bm{w}_{\text{e}}^{(j)}$
\end{enumerate}

The intelligent model weight decryption process is performed on the local device, which is the reverse operation of the encryption process. The decryption process executes the de-diffusion operations first to recover values of weight elements. Then, de-permute operations are performed to modify the positions of weight elements by $n_{\text{p}}$ rounds. The $1$-round de-diffussion operation and $n_{\text{p}}$-round de-permute operations are repeated $n_{\text{d}}$ rounds to obtain the decrypted intelligent model weight $\bm{w}^{(j)}$. To protect model IP in the interference process, intelligent model weights are not decrypted until they are used. Since the limitation of on-chip memory, intelligent model weights of all layers are usually stored on the off-chip memory. When performing an AI model, the weights are loaded on the chip layer by layer. The execution of the layer $j \in \{1,2, \cdots, N_{\text{w}}\}$ includes the following steps: 1) Load the encrypted weight $\bm{w}_{\text{e}}^{(j)}$ and the input of this layer $\bm{x}^{(j)}$ on the chip. 2) Decrypt the encrypted weight $\bm{w}_{\text{e}}^{(j)}$ and obtain the $\bm{w}^{(j)}$. 3) Perform the inference operations at the layer $j$ of the intelligent model, i.e $\bm{x}^{(j+1)}=f_j(\bm{w}^{(j)}\bm{x}^{(j)})$, where $f_j(\cdot)$ is the activation function of layer $j$. 4) Store the output $\bm{x}^{(j+1)}$ of layer $j$ to the off-chip memory, which is also the input of the next layer.

\subsection{Permute and Diffusion Operations in AI Model Protection}

The permute and diffusion operations in the AI model encryption/decryption are executed based on the secret keys. Suppose the size of the secret key is $2n$, which is denoted as $\bm{k}=(k1,k2,\cdots, k_n, k_{n+1}, k_{n+2}, \cdots, k_{2n})$. The secret key $\bm{k}$ is divided into two parts: $\bm{k}_{\text{p}}=(k1,k2,\cdots, k_n)$ and $\bm{k}_{\text{d}}=(k_{n+1}, k_{n+2}, \cdots, k_{2n})$, which are utilized for permute and diffusion operations respectively.

\subsubsection{Permute Operations in Model Encryption/Decryption}
Based on the PUF-enabled secret key $\bm{k}_{\text{p}}$, chaos theory is utilized to permute positions of elements of intelligent model weights. The Chaos system has the capability to generate random sequences in a deterministic system and the random sequence has only a relationship to the initial states of the system.

The logistic chaos function is a classic and effective mapping function \cite{wu2014discrete}, which can be adopted to generate a random permute sequence, which is expressed as follows:
\begin{equation}
s_{i+1} = \lambda s_i (1-s_i), \quad s_i \in [0,1],
\label{log}
\end{equation}
where $s_i$ is the bit $i$ of the random pertume sequence. When $\lambda \in (3.59,3.99)$, the Logistic mapping function in (\ref{log}) will produce a complex chaotic sequence. Since the chaos random sequence generated in (\ref{log}) has related to initial states $\lambda$ and $s_0$ merely, the PUF-based secret key $\bm{k}_{\text{p}}$ is applied to determine these initial states as follows:
\begin{equation}
  \begin{split}
  s_0 &= \big({k(1)\oplus k(2) \oplus \cdots \oplus k(b_{\text{p}})}\big) / {b_{\text{p}}}\\
  \label{logsta}
  \end{split}
\end{equation}
\begin{equation}
  \begin{split}
  \lambda & = 3.6 + 0.2\times\big({k(b_{\text{p}}\!\!+\!\!1)\oplus k(b_{\text{p}}\!\!+\!\!2) \oplus \cdots \oplus k(2b_{\text{p}})}\big) {b_{\text{p}}}  
  \label{logstb}
  \end{split}
\end{equation}
where $k(1), k(2), \cdots, k(b_{\text{p}}),k(b_{\text{p}}+1), \cdots , k(2b_{\text{p}})$ are $2b_{\text{p}}$ secret key bits selected from $\bm{k}_{\text{p}}$ and $2b_{\text{p}}\leq n$. 

To permute the intelligent model weight $\bm{w}^{(j)} \in \mathcal{R}^{L_{j} \times L_{j-1}}$ for $j=1,2, \cdots, N_{\text{w}}$, the logistic chaos system runs $T_{\text{pre}}$ rounds to get into chaos first, and then, executes $L_{j} \times L_{j-1}$ times to the random chaos sequence $\bm{s}=(s_1,s_2,\cdots, s_{L_{j} \times L_{j-1}})$. After that, $\bm{s}$ is sorted in ascending/descending order, and the sorted random sequence $\tilde{\bm{s}}$ is obtained. Next, the converted weight $\bm{w}'^{(j)}$ is permuted based on position exchange between $\bm{s}$ and $\tilde{\bm{s}}$. The permute operations are executed $n_{\text{p}}$ rounds and obtian the permuted weight $\bm{w}''^{(j)}$.

The de-permute operation is reversed to the permute operation. The local device select identical bits from the secret key $\bm{k}_{\text{p}}$ and generates $s_0$ and $\lambda$ based on (\ref{logsta}) and (\ref{logstb}). Based on the Logistic chaos mapping function, the random sequence $\bm{s}$ and the sorted random sequence $\tilde{\bm{s}}$ are generated. Then, the local devices de-permute the intelligent weight according to the relationship between $\bm{s}$ and $\tilde{\bm{s}}$.

\subsubsection{Diffusion Operations in Model Encryption/Decryption}

The diffusion module is connected to the permute module and aims to transform values of AI model weights. The proposed diffusion module is based on the nonlinear chaos algorithms (NAC) and the PUF-enabled secret key $\bm{k}_{\text{d}}$. The NAC is utilized to generate diffusion chaos sequence, which is expressed as follows \cite{hussain2012novel}:
\begin{equation}
   \begin{split}
      s_{i+1}^{\text{N}} &= \gamma \tan(\alpha s_i^{\text{N}})(1-s_i^{\text{N}})^{\beta}, \quad s_i^{\text{N}} \in [0,1] 
   \label{nac}
   \end{split}
\end{equation}
where  
\begin{equation}
\gamma = (1-\beta^{-4}) \cot(\frac{\alpha}{1+\beta})(1+\frac{1}{\beta})^{\beta}.
\end{equation}
When $\alpha \in (1,1.4]$ and $\beta \in [5,43]$, the NCA in (\ref{nac}) is chaotic. The PUF-enabled secret key $\bm{k}_{\text{d}}$ is adopted to determine the initial state $s_{0}^{\text{N}}$ and parameters $\alpha$ and $\beta$ of the NCA, which are listed as follows:
\begin{equation}
  \begin{split}
  s_{0}^{\text{N}} &= \big({k^{\text{N}}(1)\oplus k^{\text{N}}(2) \oplus \cdots \oplus k^{\text{N}}(b_{\text{d}})}\big)/{b_{\text{d}}}\\
  \label{nacsta}
  \end{split}
\end{equation}
\begin{equation}
  \begin{split}
  \alpha & = 1.1 +0.35 \times \big({k^{\text{N}}(b_{\text{d}}+1)\oplus k^{\text{N}}(b_{\text{d}}+2) \oplus \cdots \oplus k^{\text{N}}(2b_{\text{d}})}\big) / {b_{\text{d}}} 
  \label{nacalp}
  \end{split}
\end{equation}
\begin{equation}
  \begin{split}
  \beta & = 6 + 35\times  \big({k^{\text{N}}(2b_{\text{d}}\!\!+\!\!1)\oplus k^{\text{N}}(2b_{\text{d}}\!\!+\!\!2) \oplus \cdots \oplus k^{\text{N}}(3b_{\text{d}})}\big)/ {b_{\text{p}}} 
  \label{nacbet}
  \end{split}
\end{equation}
where $k^{\text{N}}(1), k^{\text{N}}(2), \cdots, k^{\text{N}}(b_{\text{p}}), \cdots , k^{\text{N}}(2b_{\text{p}}), \cdots, k^{\text{N}}(3b_{\text{p}})$ are $3b_{\text{p}}$ secret key bits selected from $\bm{k}_{\text{b}}$, which satisfies $3b_{\text{b}}\leq n$. 

The NCA mapping function executes $T_{\text{pre}}$ times to get into a chaotic state. Then $L_j \times L_{j-1}$ rounds are performed to generate the diffusion chaos sequence $\bm{s}^{\text{N}}=(s_{1}^{\text{N}},s_{2}^{\text{N}}, \cdots, s_{L_j \times L_{j-1}}^{\text{N}})$. The permuted weight $\bm{w}''^{(j)}$ is diffused as follows:
\begin{equation}
  \begin{split}
  \bm{w}'''^{(j)} &= \begin{cases}
                                \bm{w}''^{(j)} + \bm{s}^{\text{N}}, \quad  \bm{s}^{\text{N}}<0.5 \\
                                 \bm{w}''^{(j)} - \bm{s}^{\text{N}}, \quad  \bm{s}^{\text{N}}\geq 0.5
                               \end{cases}
  \end{split}
\end{equation}
%where $\circ$ is the Hadamard product. 
The intelligent model provider runs the $n_{\text{p}}$-round permute operations and $1$-round diffusion operation for $n_{\text{d}}$ times. Then, the permuted and diffused model weight is recovered to multiple-dimensional encrypted weight $\bm{w}_{\text{e}}^{(j)}$. The de-diffusion is the reverse operation of the diffusion. In the process of de-diffusion, identical secret key bits are selected from $\bm{k}_{\text{b}}$ to generate identical diffusion chaos sequence $\bm{s}^{\text{N}}$ based on the NCA mapping function. According to the diffusion chaos sequence $\bm{s}^{\text{N}}$, de-diffusion operations are executed.

\section{Experimental Evaluation}
\label{sec:eva}
The effectiveness of the proposed AI model protection approach is verified by a series of experiments in this section. We discuss the environment settings first, and then, experimental results are presented.

\subsection{Environment Settings}

The experiments are implemented based on the TensorFlow framework. Several commonly used deep neural networks (DNN) are adopted in the evaluation, i.e. multilayer perceptron (MLP) and convolutional neural networks (CNN). The CNN used in this section includes a simple CNN model and a complex CNN model VGG-16. These intelligent models are evaluated on the Mnist, Fashion-mnist, Cifar10, and Cifar 100 datasets for image classification. The detailed information of the evaluated intelligent models and datasets are presented in Table \ref{settings}. The intelligent model encryptions are implemented on the convolutional layers and fully connected layers.
\setlength{\extrarowheight}{0.15mm}
\begin{table}[htbp]
\renewcommand{\arraystretch}{1.1}
 %if using array.sty, it might be a good idea to tweak the value of
 %\extrarowheight as needed to properly center the text within the cells
\caption{Evaluated intelligent models and datasets}
\label{settings}
\centering
% Some packages, such as MDW tools, offer better commands for making tables
% than the plain LaTeX2e tabular which is used here.
\begin{threeparttable}
\begin{tabular}{|c|c|c|}
    \hline
    {\textbf{Types}}&{\textbf{Models/datasets}} & {\textbf{Description}}\\
    \hline
    \hline
     \multirow {3}{*}{Models} & MLP & 4 ReLU, 5FC \\
      			                                              & Simple CNN & 3 C, 2 MP, 4 ReLU, 2 FC\\
                                                           & VGG-16& 13 C, 4 MP, 15 ReLU, 3 FC\\
                                                            
     \hline
     \multirow {6}{*}{ Datasets} & \multirow {3}{*}{Mnist/Fashion-Mnist} & Input size: $28 \times 28$\\               
                                              &                                      & Classification: $10$\\
                                              &                                     & Train/test size: $60000/10000$\\
                
               \cline{2-3}
      	 	      &  \multirow {3}{*}{Cifar-10/Cifar-100} & Input size: $28 \times 28 \times 3$\\               
                                              &                                     & Classification: $10/100$\\
                                              &                                     & Train/test size: $50000/10000$\\
               
    \hline
\end{tabular} 
\begin{tablenotes}
\footnotesize
\item[] Note: FC: full connected layer, C: convolutional layer, MP: maximum pooling layer.
\end{tablenotes}
\end{threeparttable}
\end{table}

\subsection{Experimental Results}

The experimental results include the following four parts, i.e. performance of intelligent models encryption, anti-fine tune attack ability, the influence of encryption parameters, and applications on complex models.
\setlength{\extrarowheight}{0.15mm}
\begin{table}[htbp]
\renewcommand{\arraystretch}{1.1}
\caption{accuracy of CNN models}
\label{origionalacc}
\centering
\begin{tabular}{|c|c|c|}
    \hline
    {\textbf{Models}}&{\textbf{Datasets}} & {\textbf{Origional accuracy}} \\
    \hline
    \hline
     MLP & Mnist &  98.49\% \\
     MLP & Fashion-mnist & 89.64\%\\
     CNN & Mnist &  99.32\% \\
     CNN & Fashion-mnist & 91.57\%\\
   \hline
\end{tabular} 
\end{table}

\subsubsection{Performance of Intelligent Models Encryption}
\begin{figure}[t]
\centering
\includegraphics[width=3.5in]{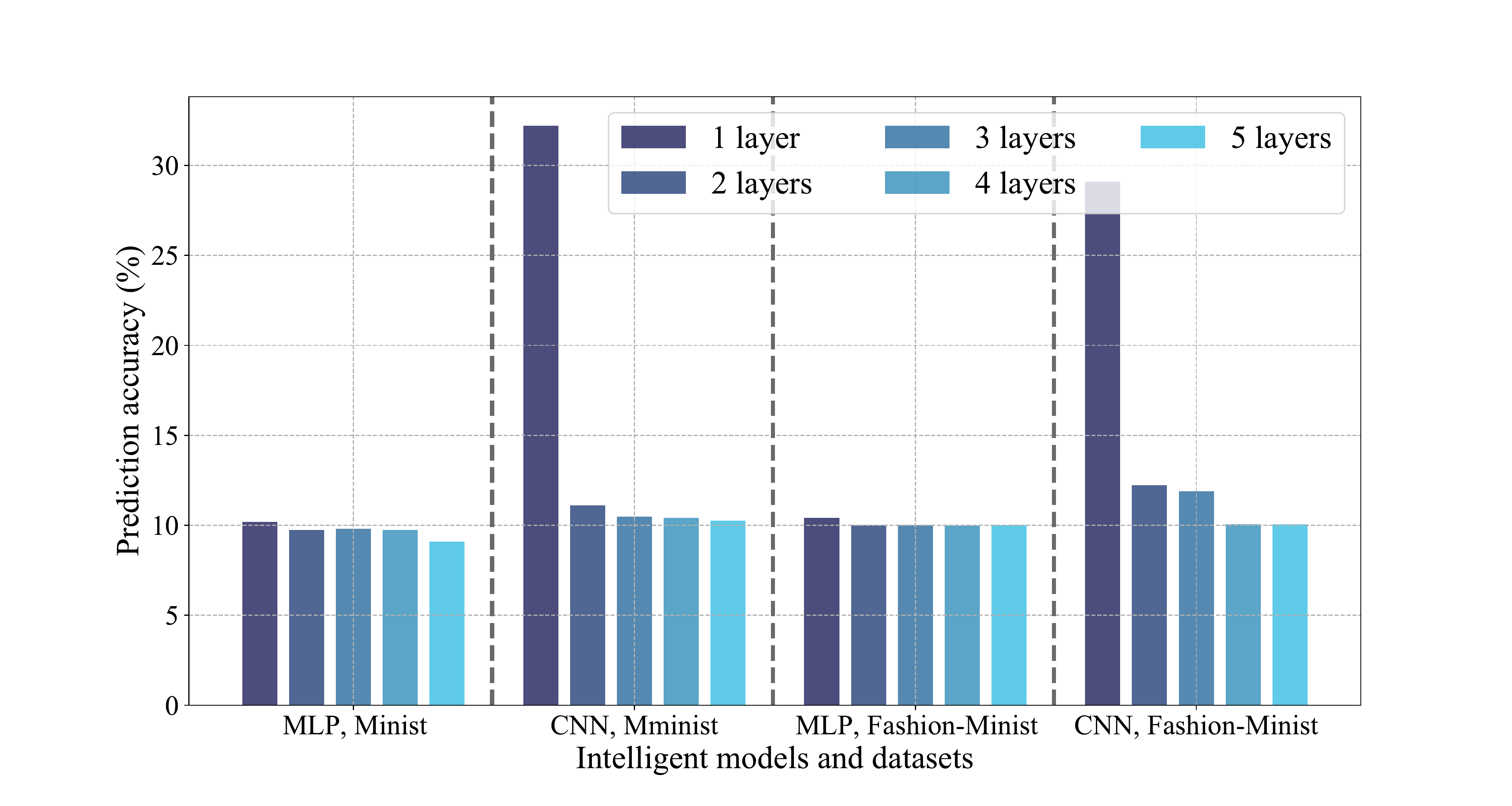}
\centering
\caption{Prediction accuracy of models encrypted on a single layer.}
\label{fig:multi_acc}
\end{figure}

\begin{figure}[t]
\centering
\includegraphics[width=3.5in]{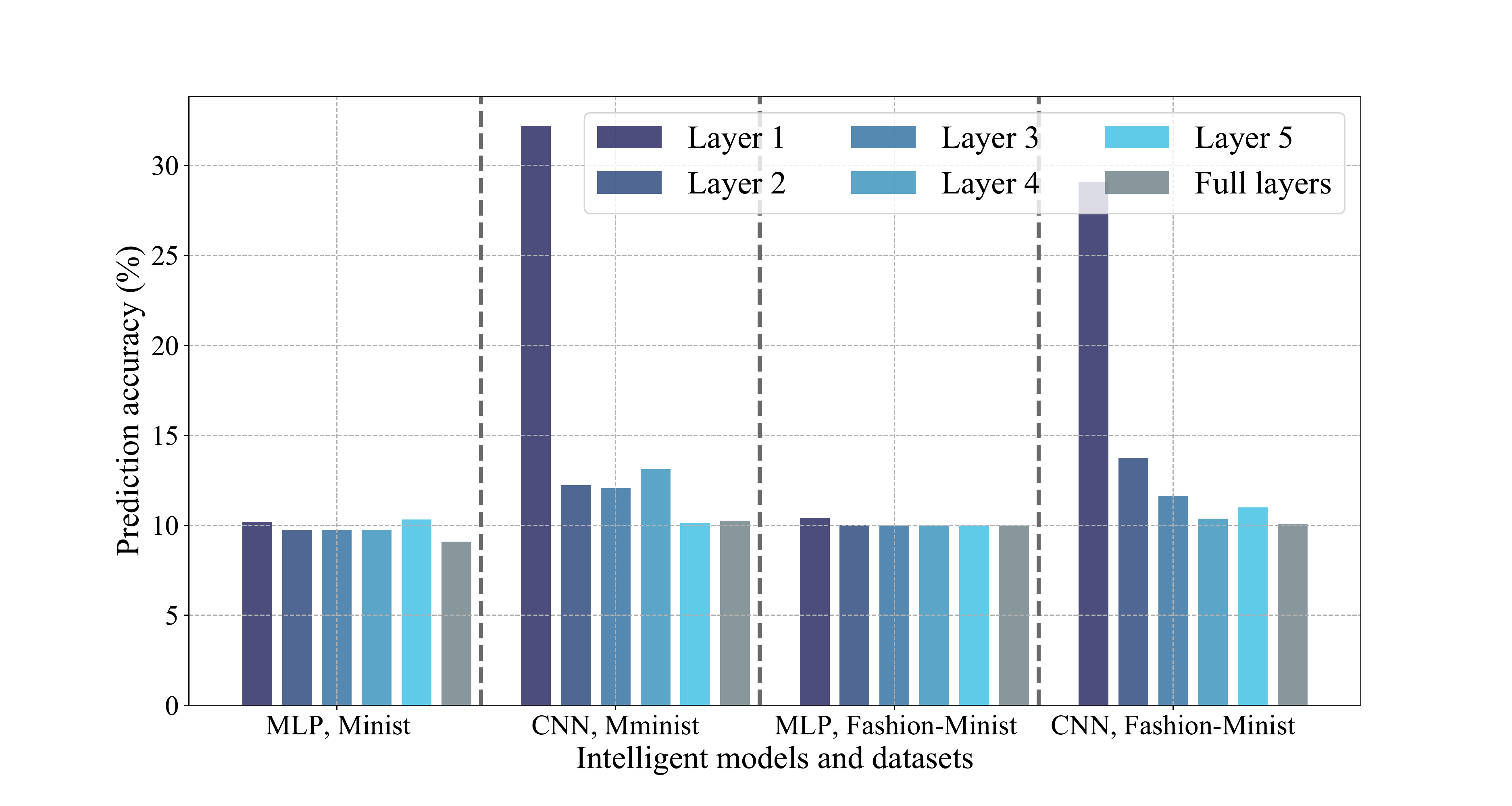}
\centering
\caption{Prediction accuracy of the specific-layer encrypted models.}
\label{fig:single_acc}
\end{figure}

The performance of the proposed intelligent models IP protection is shown in Fig. \ref{fig:multi_acc} and Fig. \ref{fig:single_acc}, which investigate the prediction accuracy of encrypted MLP and simple CNN modes on Mnist and Fashion-mnist datasets. The original accuracy of the MLP and simple CNN models on Mnist and Fashion-mnist datasets are shown in Tab. \ref{origionalacc}. 

In Fig. \ref{fig:multi_acc}, different numbers of layers in intelligent models are encrypted. Specifically, the first 1, 2, 3, 4, and 5 layers of the MLP and simple CNN models are investigated respectively, and then, these encrypted models are evaluated on the test datasets. The proposed intelligent model protection mechanism works well on the MLP models. The prediction accuracy of encrypted MLP modes drops to about $10\%$, where the adversary cannot get a better prediction than a random guess. For the simple CNN model, only encrypting the first layer of the model is not enough for protection since the accuracy only drops to $32.21\%$ and $29.10\%$ on Mnist and Fashion-mnist datasets respectively. With the increasing of encrypted layers, the prediction accuracy drops to about $10\%$ on both test datasets. In Fig. \ref{fig:single_acc}, the prediction accuracy of MLP and simple CNN models are encrypted on the single layer. From the results of Fig. \ref{fig:single_acc}, we can obtain that the protection performance of the single-layer encrypted model is comparable to that of the multiple-layer encrypted model except for only encrypting the first layer of simple CNN.

\subsubsection{Anti-Fine Tuning Attack Ability}
\begin{figure*}[t]
\centering
\subfigure[]{
\begin{minipage}[t]{0.25\linewidth}
\centering
\includegraphics[width=1.85in]{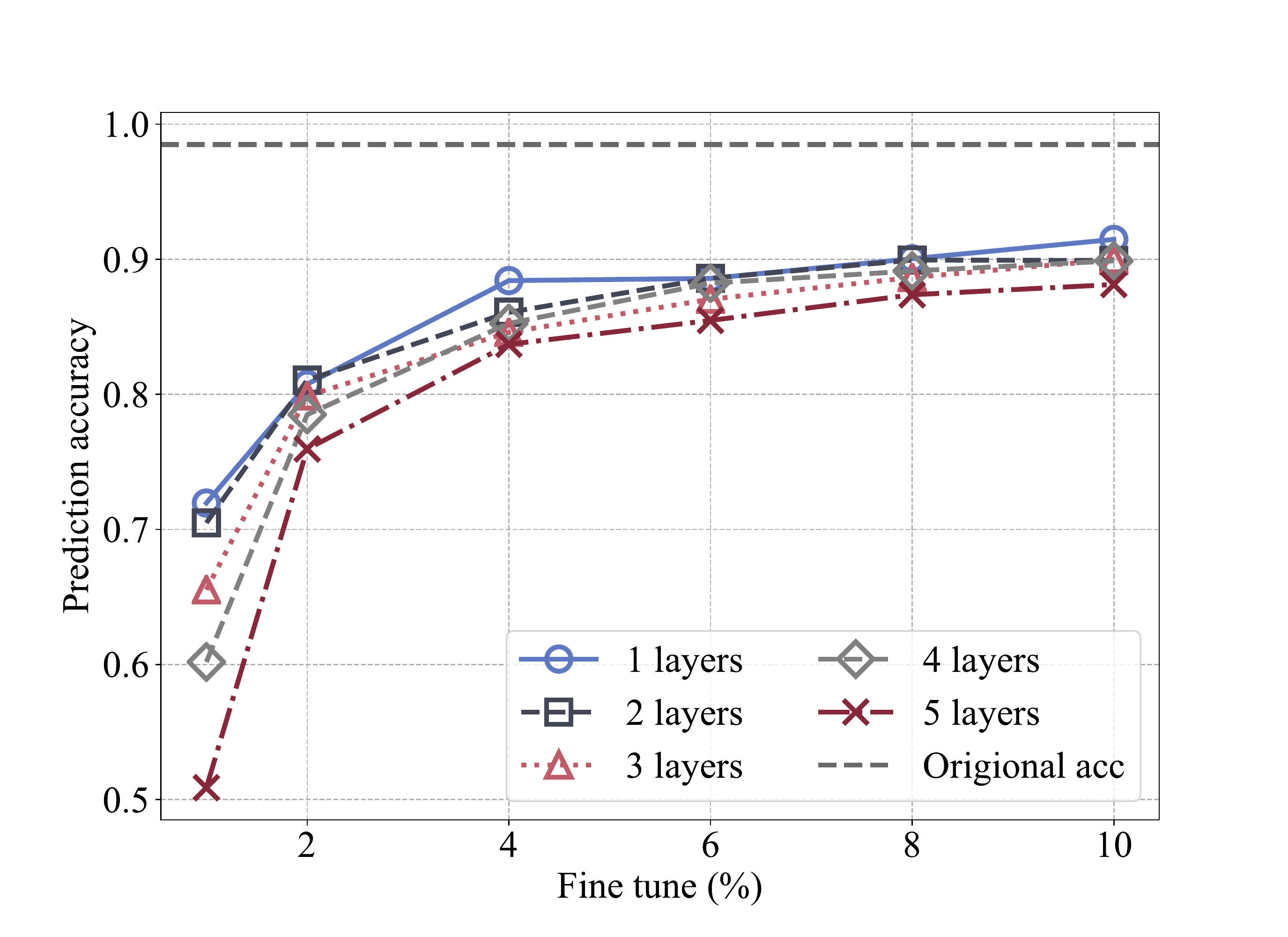}
%\caption{fig1}
\end{minipage}%
}%
\subfigure[]{
\begin{minipage}[t]{0.25\linewidth}
\centering
\includegraphics[width=1.85in]{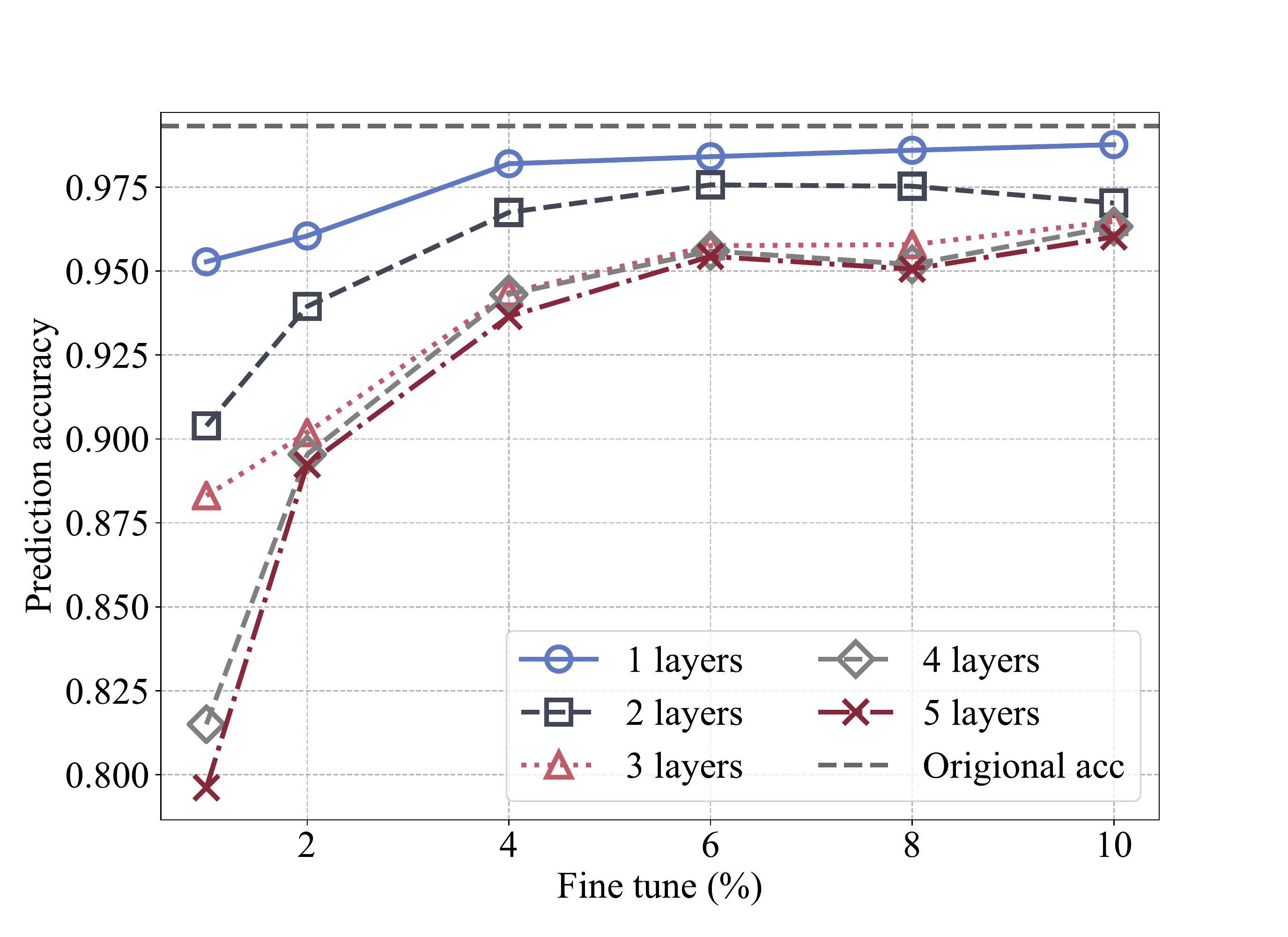}
%\caption{fig2}
\end{minipage}%
}%
\subfigure[]{
\begin{minipage}[t]{0.25\linewidth}
\centering
\includegraphics[width=1.85in]{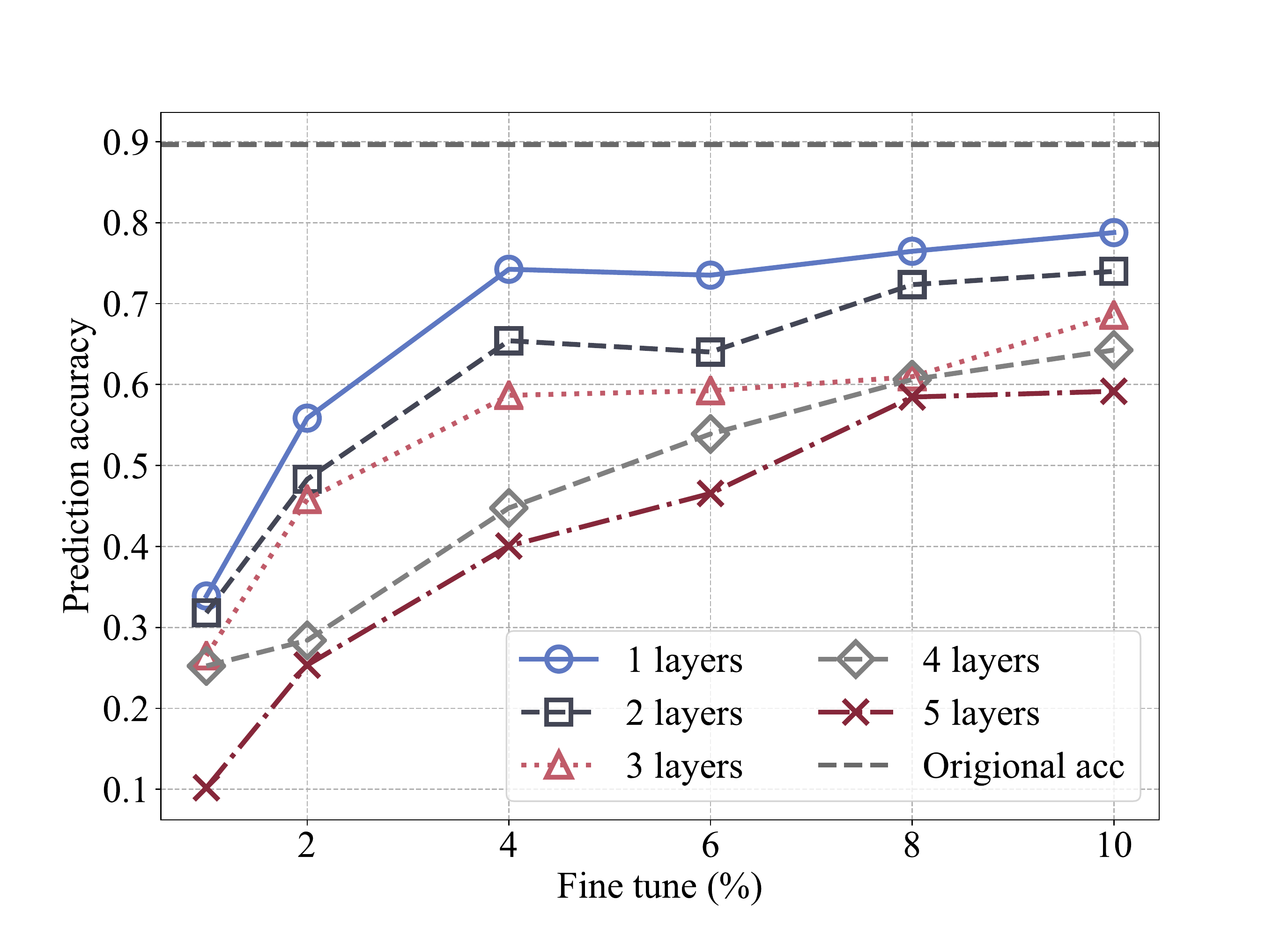}
%\caption{fig1}
\end{minipage}%
}%
\subfigure[]{
\begin{minipage}[t]{0.25\linewidth}
\centering
\includegraphics[width=1.85in]{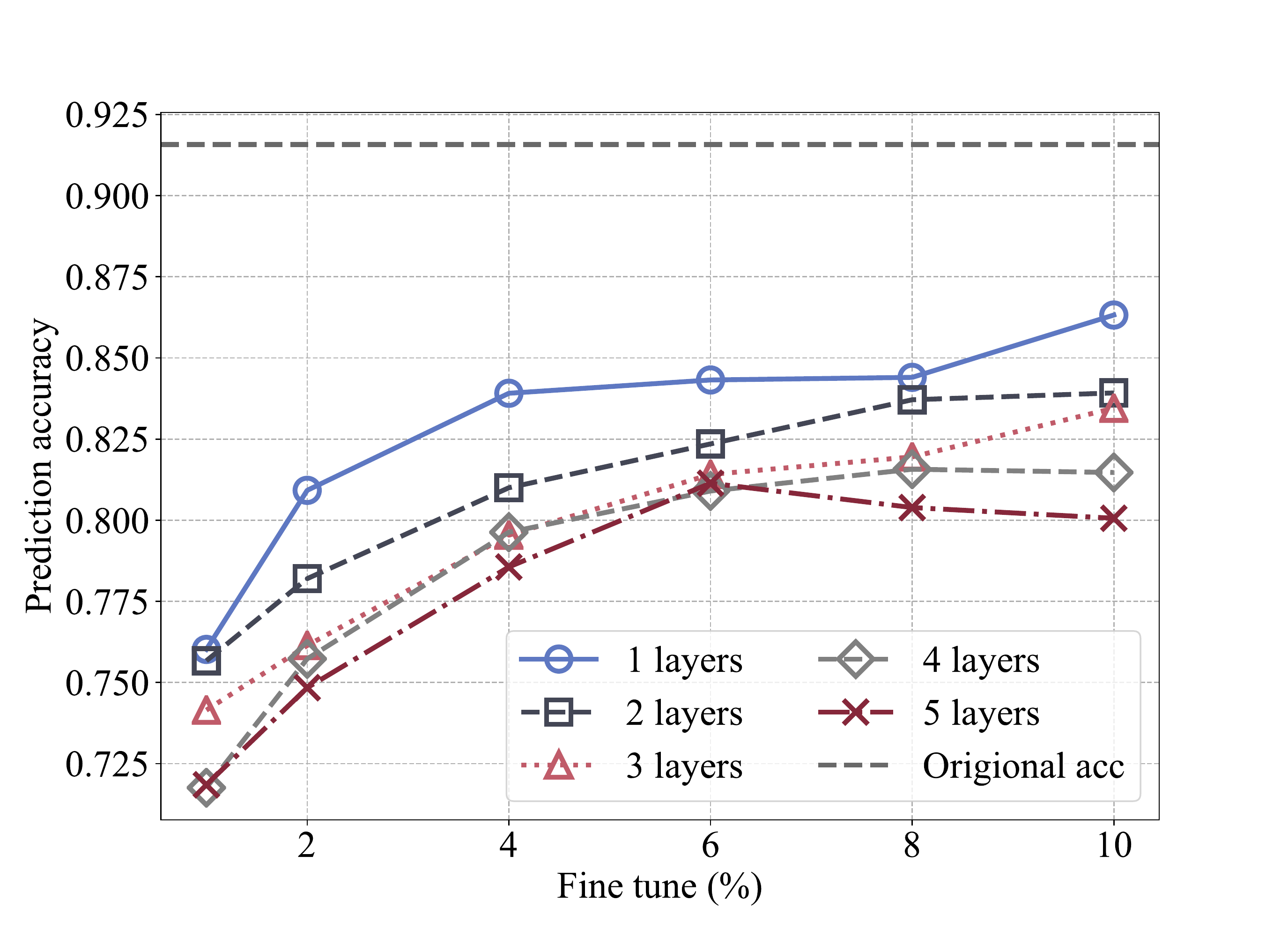}
%\caption{fig2}
\end{minipage}%
}%
\caption{Fine-tuning attacks on intelligent models with various fractions of the Mnist and Fashion-mnist training dataset: a) MLP on Mnist; b) simple CNN on Mnist; c) MLP on Fashion-mnist; d) simple CNN on Fashion-mnist.}
\label{fig:finetune}
\end{figure*}
\setlength{\extrarowheight}{0.15mm}
\begin{table*}[htbp]
\renewcommand{\arraystretch}{1.1}
\caption{Fine-tuning attacks on CNN with single-layer encryption}
\label{tab:singlefinetune}
\centering
\begin{tabular}{|c|c|c|cc|cccccc|}%|c|c|c|c|c|c|c|c|c|c|
    %\toprule
    \hline
    \multirow {2.5}{*}{\textbf{Dataset}}&\multirow {2.5}{*}{\textbf{Layer}}& \multirow {2.5}{*}{\makecell[c]{\textbf{No. of encrypted}\\ \textbf{parameters}} }& \multicolumn{2}{c|}{\textbf{Model Encryption} }& \multicolumn{6}{c|}{\textbf{Fine-tuning accuracy}}\\
    \cline{4-11}
    %\cmidrule{4-5} \cmidrule{6-11}
    & & & {Accuracy} & {Drop} & 1\% &2\% &4\% & 6\%&8\%& 10\%\\
    \hline
    \hline
    \multirow {5}{*}{Mnist}
    		 &Conv 1  &   $320$     & 32.21\%& 67.11 \% &93.87\% & 96.15\% & 96.80\% & 97.73\% & 98.10\% & 98.78\% \\
		     &Conv 2  &   $18496$  & 12.21\%& 87.11 \% & 92.18\% & 93.75\% & 96.36\% & 97.07\% & 97.52\% & 97.51\% \\
		     &Conv 3  &   $73856$  & 12.08\%& 87.24 \%& 87.26\% & 91.17\% & 93.59\% & 95.63\% & 96.32\% & 96.63\% \\
		     &Dense 1  & $147584$ & 13.12\%& 86.20 \% & 85.19\% & 90.44\% & 93.30\% & 94.78\% & 95.84\% & 96.22\% \\
     		 &Dense 2  & $1290$    & 10.12\%& 89.20 \%& 81.19\% & 87.99\% & 92.66\% & 94.31\% & 95.67\% & 96.04\% \\
     \hline
    \multirow {5}{*}{Fashion-mnist}
    		 &Conv 1  &   $320$     & 29.10\%& 62.47 \%& 79.85\% & 82.06\% & 82.75\% & 83.99\% & 85.36\% & 86.24\% \\
		     &Conv 2  &   $18496$  & 13.75\%& 77.82 \%& 74.36\% & 77.16\% & 79.27\% & 81.21\% & 83.84\% & 83.50\% \\
		     &Conv 3  &   $73856$  & 11.64\%& 79.93 \%& 72.72\% & 74.59\% & 78.46\% & 80.24\% & 81.49\% & 82.06\% \\
		     &Dense 1 & $147584$ & 10.38\%& 81.19 \% & 70.89\% & 74.92\% & 77.75\% & 80.69\% & 80.62\% & 82.50\% \\
     		 &Dense 2 & $1290$    & 11.00\%& 80.57 \% & 72.45\% & 74.69\% & 76.74\% & 80.25\% & 80.87\% & 82.15\% \\
     \hline
     %\bottomrule
\end{tabular} 
\end{table*}
To evaluate the effectiveness of the proposed intelligent model protection mechanism, we investigate its ability to resist the fine-tuning attack \cite{liu2018fine}. In the fine-tuning attack, suppose the adversaries have the encrypted intelligent models obtained through illegal ways and a small fraction of the training dataset used to fine-tune the encrypted models. To analyze the resistance ability of the proposed intelligent model protection mechanism to fine-tune attacks, we investigate the performance of encrypted models which are fine-tuned by the adversaries. In Fig. \ref{fig:finetune}, we investigate the fine-tuning attacks on two different intelligent models (i.e. MLP and simple CNN) with various fractions (i.e. $1\%$,$2\%$,$4\%$,$6\%$,$8\%$, and $10\%$) of the Mnist and Fashion-mnist training dataset.  Figure \ref{fig:finetune} (a) (b) (c) (d) present the fine-tuning attacks on MLP model with Mnist, CNN with Mnist, MLP with Fashion-mnist, and CNN with Fashion-mnist, respectively. From the experimental results in Fig. \ref{fig:finetune}, we can obtain that with the increasing fraction of training datasets, the adversaries achieve intelligent models with higher prediction accuracy. As the theft dataset of fine-tuning attackers up to $10\%$ of the original datasets, the prediction accuracy of MLP on Mnist, CNN on Mnist, MLP on Fashion-mnist, and CNN on Fashion-mnist are $88.13\%$, $96.02\%$, $59.17\%$, and $80.06\%$, respectively when full layers of models are encrypted. Moreover, as the number of encrypted layers increases, the prediction accuracy decreases, which means the ability to resist fine-tuning attacks increases. Although the single-layer encrypted model has a comparable protection performance to the multiple-layer encrypted models, the multiple-layer encrypted models have a higher ability to resist fine-tuning attacks.

For the specific layer of the encrypted intelligent models, the anti-fine tuning attack ability is investigated. Take simple CNN as an example, and its ability to resist fine-tuning attacks is presented in Tab. \ref{tab:singlefinetune}. From the experimental results, it can be observed that only encrypting the \textit{Conv 1} layer of the CNN model achieves lower model protection and anti-fine tuning attacks performance. This is because the \textit{Conv 1} layer has the fewest encrypted parameters and is farthest from the output of the intelligent model. Thus, encrypting the \textit{Conv 1} layer leads to the least prediction accuracy drop and fine-tuning attacks resistance.

\subsubsection{Influence of Encryption Parameters}
\begin{figure*}[t]
\centering
\subfigure[]{
\begin{minipage}[t]{0.25\linewidth}
\centering
\includegraphics[width=1.85in]{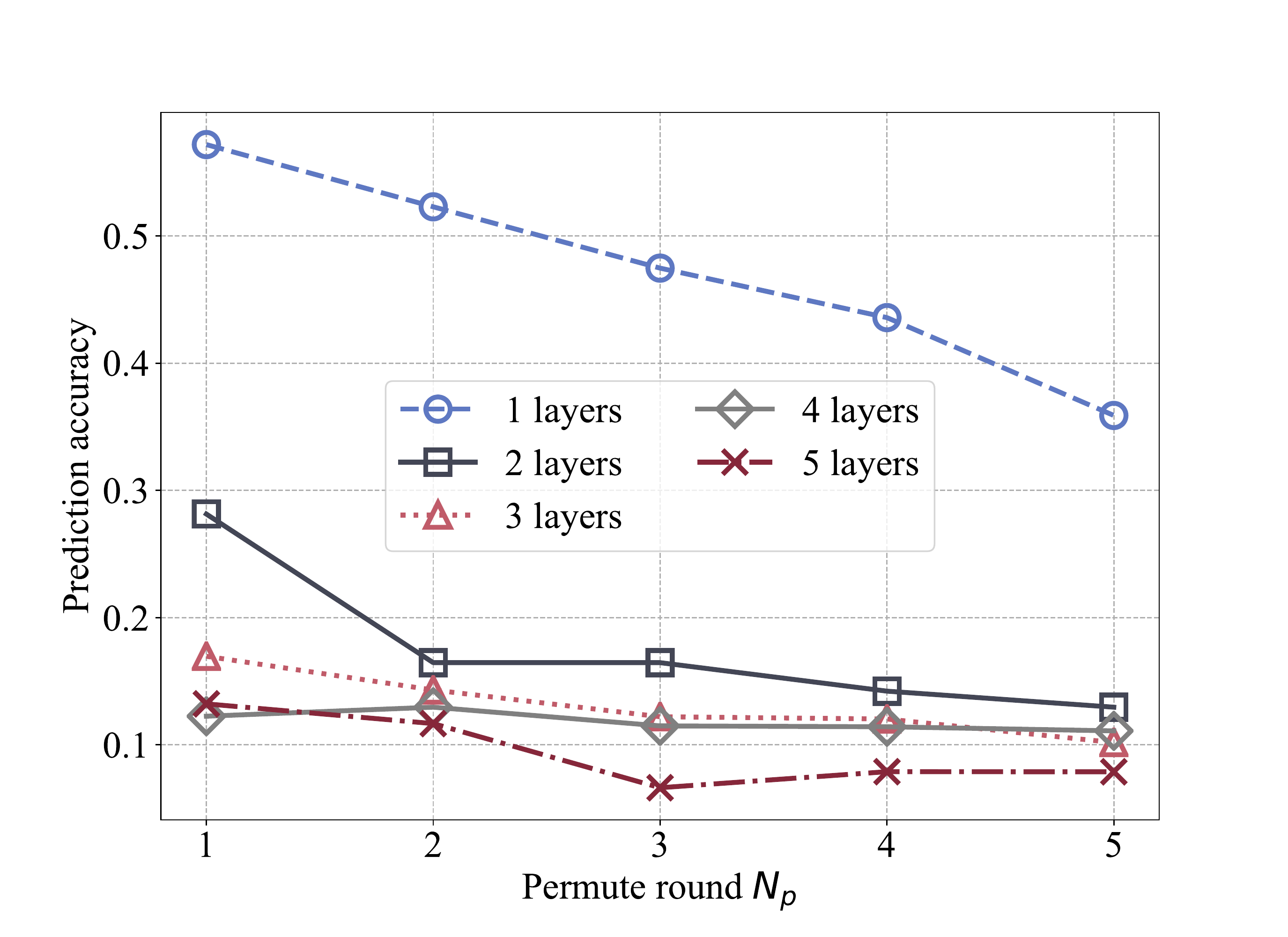}
%\caption{fig1}
\end{minipage}%
}%
\subfigure[]{
\begin{minipage}[t]{0.25\linewidth}
\centering
\includegraphics[width=1.85in]{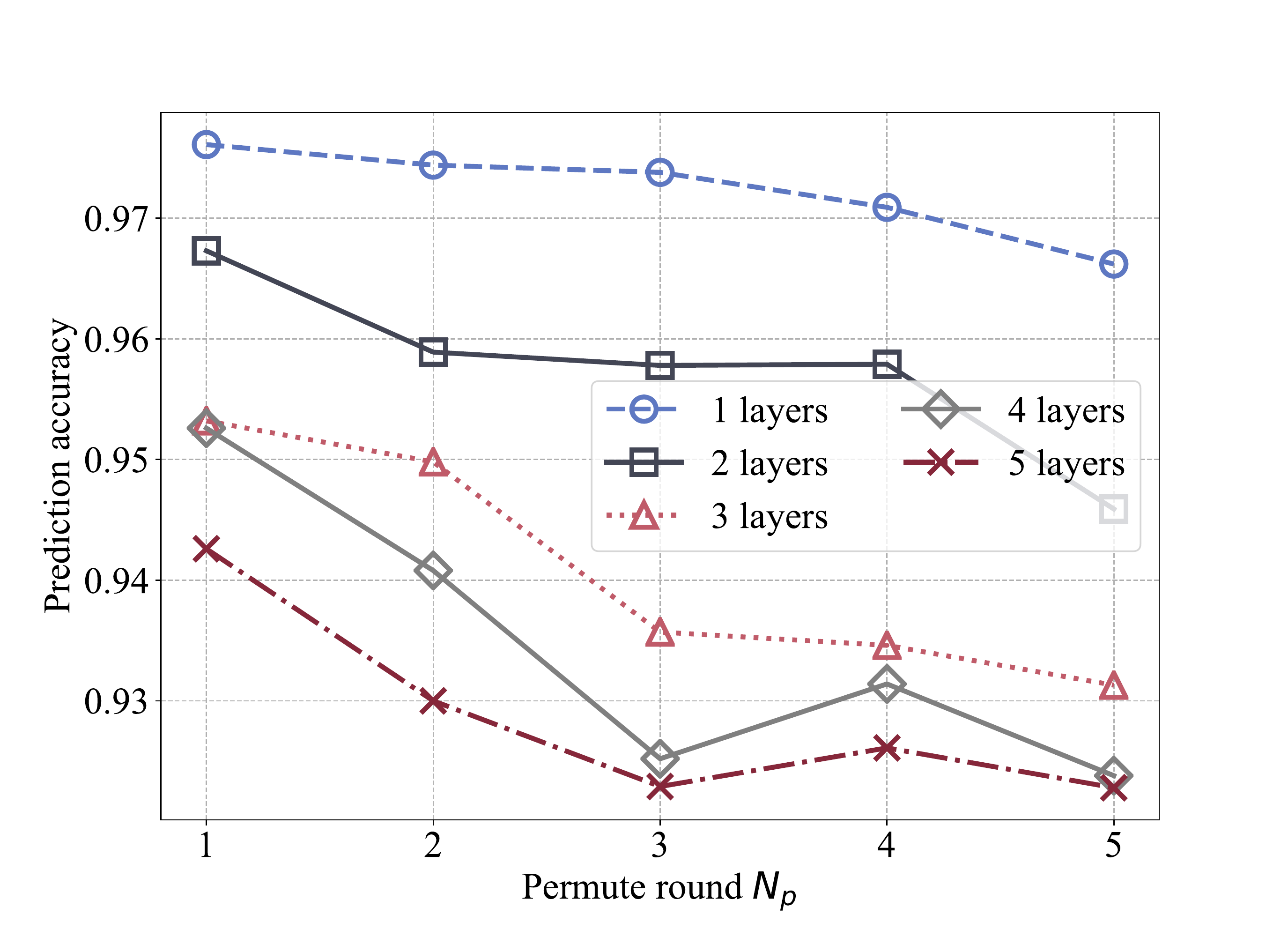}
%\caption{fig2}
\end{minipage}%
}%
\subfigure[]{
\begin{minipage}[t]{0.25\linewidth}
\centering
\includegraphics[width=1.85in]{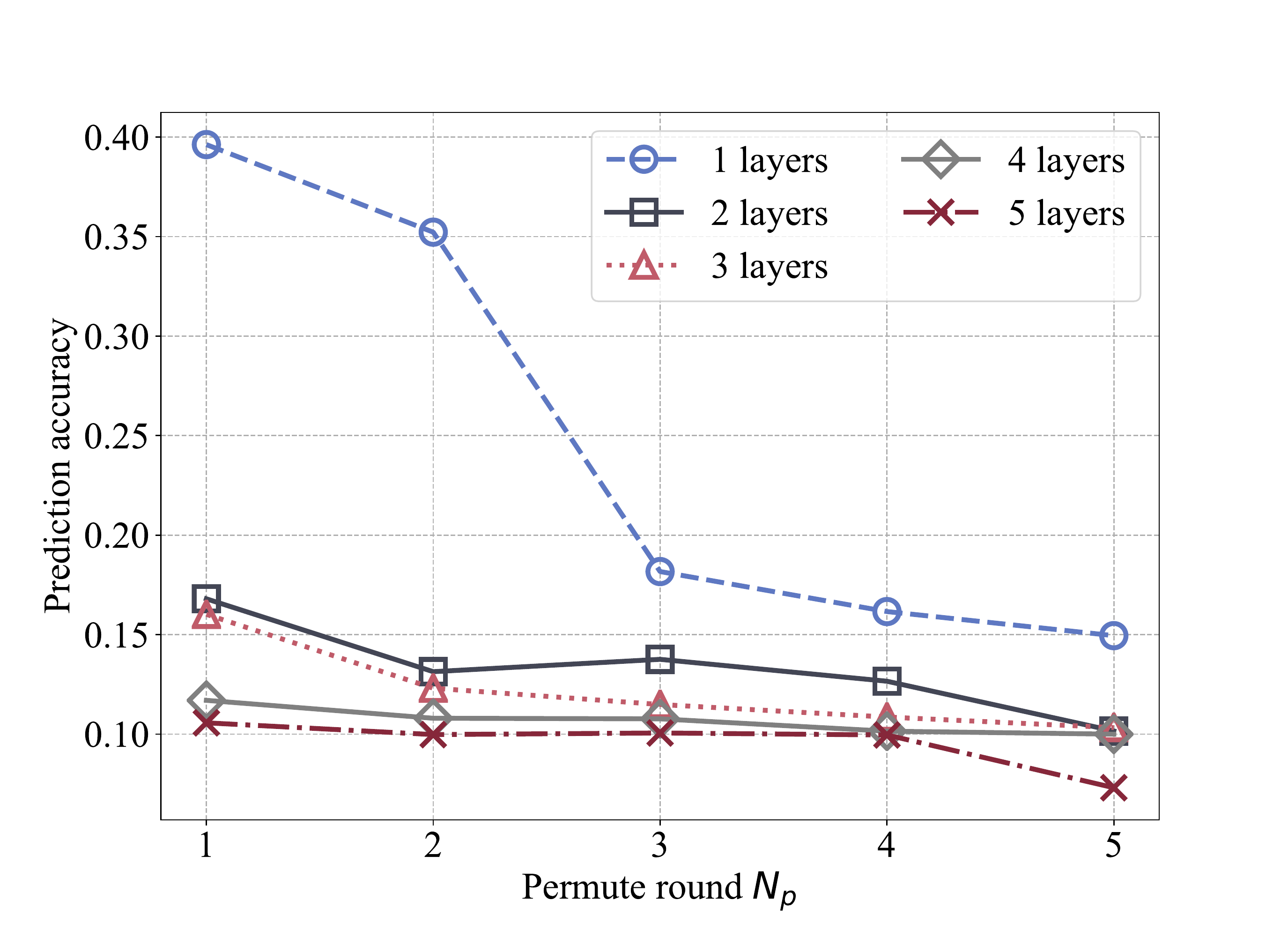}
%\caption{fig1}
\end{minipage}%
}%
\subfigure[]{
\begin{minipage}[t]{0.25\linewidth}
\centering
\includegraphics[width=1.85in]{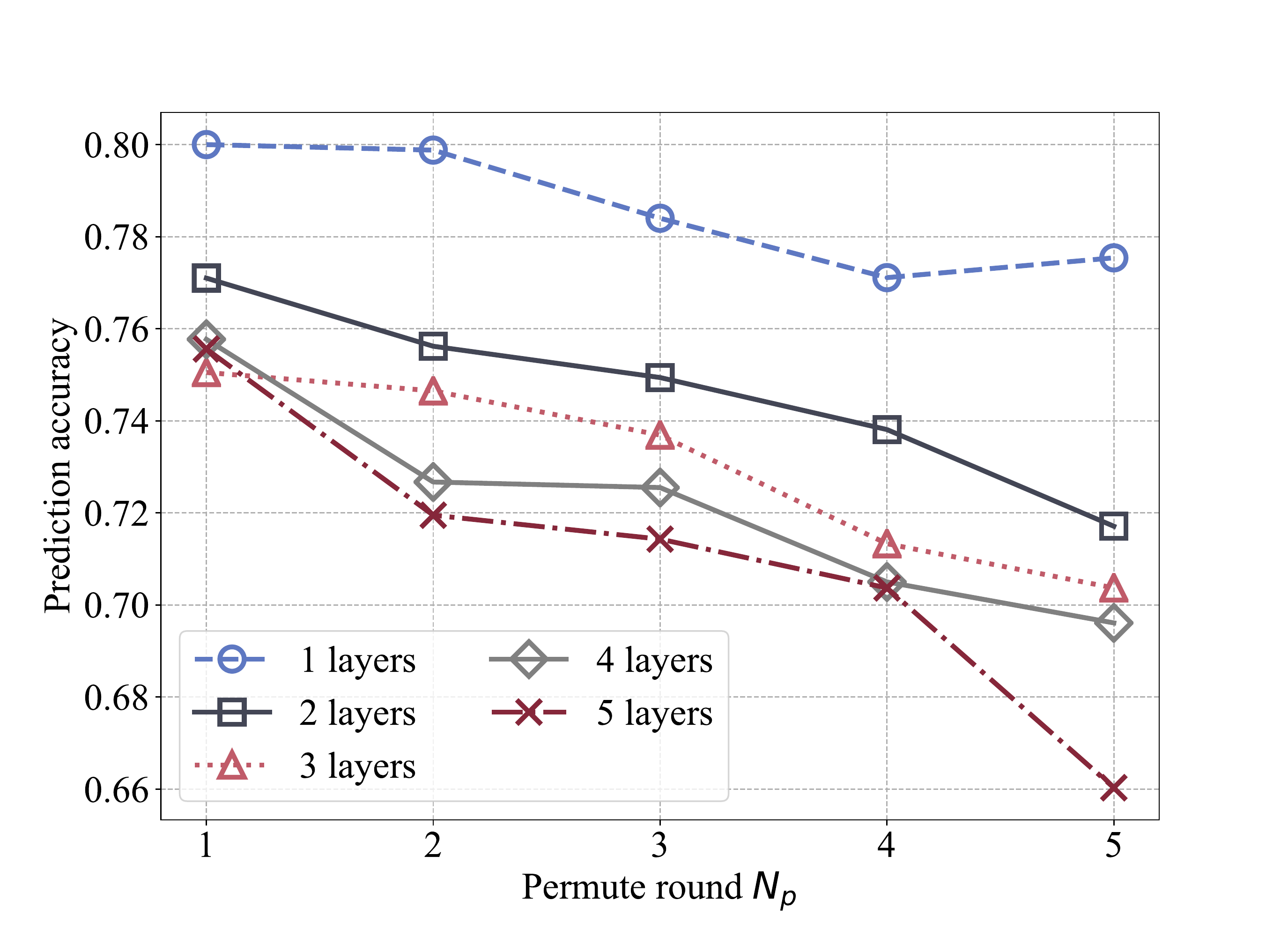}
%\caption{fig2}
\end{minipage}%
}%
\caption{Impact of the permute parameters on the simple CNN model protection: a) encrypted accuracy on Mnist; b) fine-tuning attack on the encrypted CNN on Mnist; c) encrypted accuracy on Fashion-mnist; d) fine-tuning attack on the encrypted CNN on Fashion-mnist.}
\label{fig:permute}
\end{figure*}

\begin{figure*}[t]
\centering
\subfigure[]{
\begin{minipage}[t]{0.25\linewidth}
\centering
\includegraphics[width=1.85in]{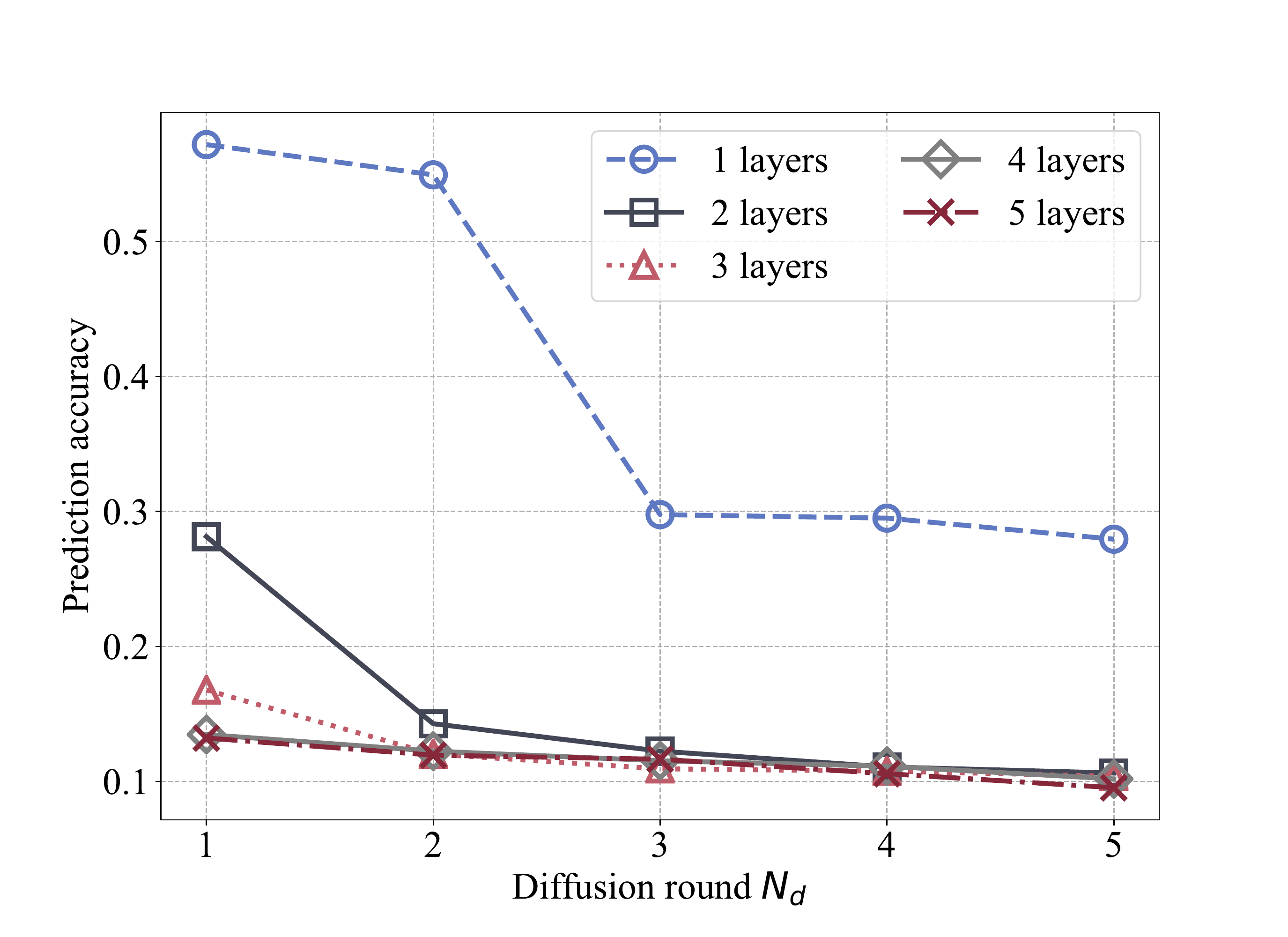}
%\caption{fig1}
\end{minipage}%
}%
\subfigure[]{
\begin{minipage}[t]{0.25\linewidth}
\centering
\includegraphics[width=1.85in]{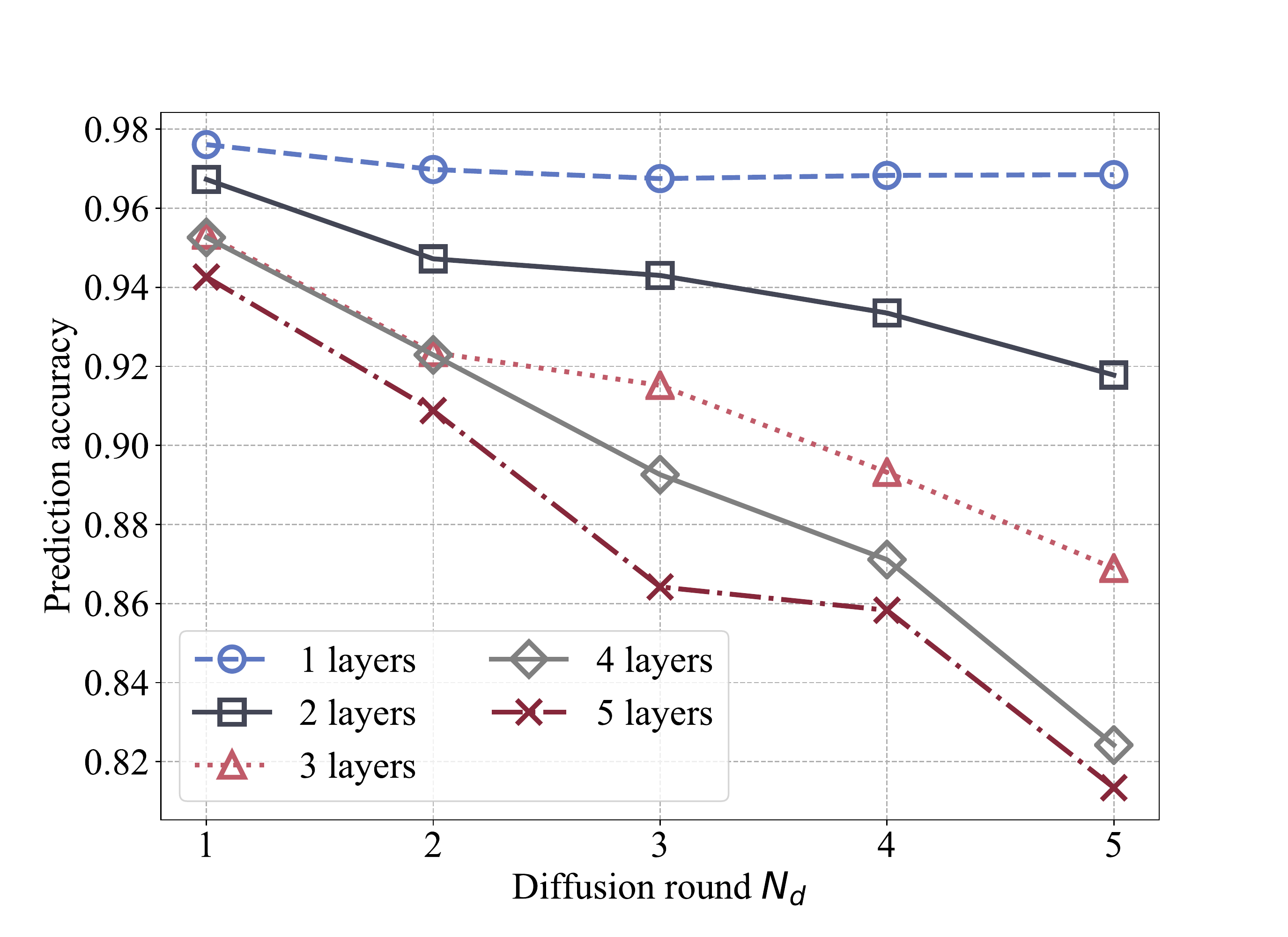}
%\caption{fig2}
\end{minipage}%
}%
\subfigure[]{
\begin{minipage}[t]{0.25\linewidth}
\centering
\includegraphics[width=1.85in]{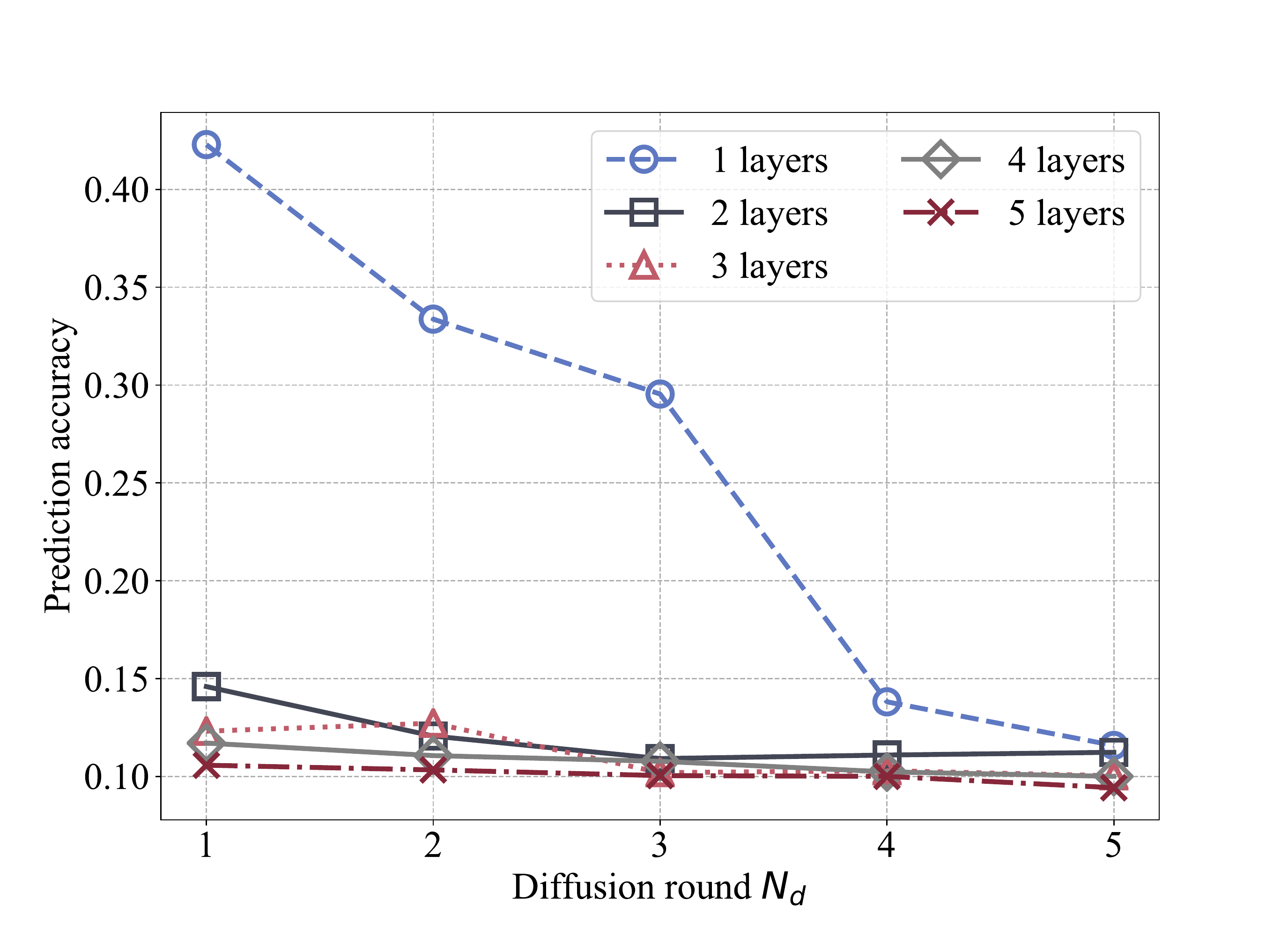}
%\caption{fig1}
\end{minipage}%
}%
\subfigure[]{
\begin{minipage}[t]{0.25\linewidth}
\centering
\includegraphics[width=1.85in]{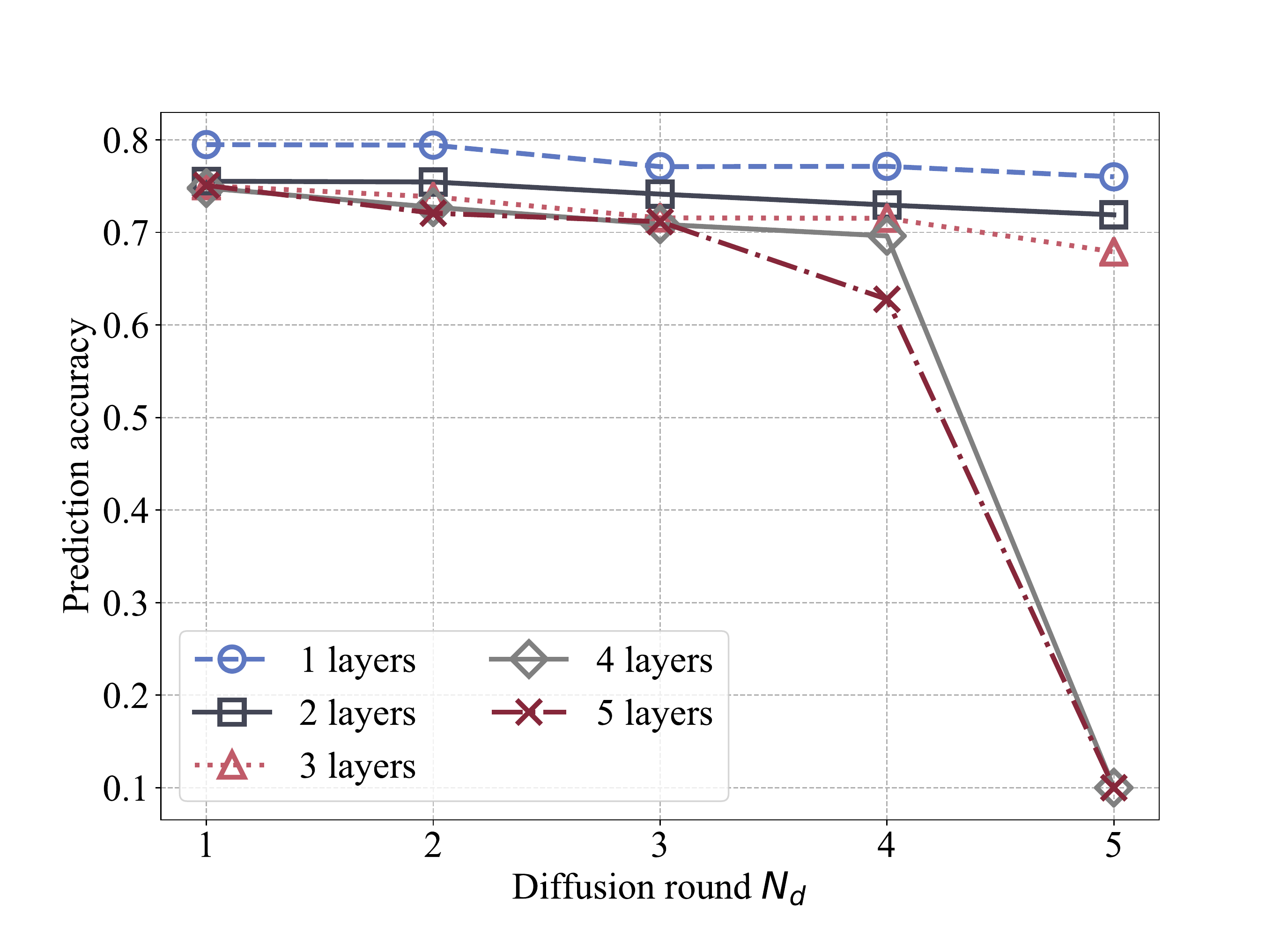}
%\caption{fig2}
\end{minipage}%
}%
\caption{Impact of the diffusion parameters on the simple CNN model protection: a) encrypted accuracy on Mnist; b) fine-tuning attack on the encrypted CNN on Mnist; c) encrypted accuracy on Fashion-mnist; d) fine-tuning attack on the encrypted CNN on Fashion-mnist.}
\label{fig:difussion}
\end{figure*}

The influence on intelligent model protection performance of encryption parameters is evaluated. We investigate the impact of the permute and diffusion parameters $n_{\text{p}}$ and $n_{\text{d}}$ on the proposed intelligent model IP protection mechanism in Fig. \ref{fig:permute} and Fig. \ref{fig:difussion}, respectively.
\setlength{\extrarowheight}{0.15mm}
\begin{table*}[htbp]
\renewcommand{\arraystretch}{1.1}
\caption{Fine-tuning attacks on CNN with single-layer encryption}
\label{tab:vgg}
\centering
\begin{tabular}{|c|c|cc|cc|cc|cc|}%|c|c|c|c|c|c|c|c|c|c|
    %\toprule
    \hline
    \multirow {2.5}{*}{\textbf{Encrypted layers}}& \multirow {2.5}{*}{\makecell[c]{\textbf{No. of encrypted}\\ \textbf{parameters}} }& \multicolumn{4}{c|}{\textbf{Cifar 10 dataset} }& \multicolumn{4}{c|}{\textbf{ Cifar 100 dataset}}\\
    \cline{3-10}
    %\cmidrule{4-5} \cmidrule{6-11}
    & &  {Accuracy} & {Drop} & 1\% Fine tune & Drop & {Accuracy} & {Drop} & 1\% Fine tune & Drop\\
    \hline
    \hline

    		 1  &   $1792$    & 11.12\%& 82.47\% &30.06\% & 63.53\% & 2.95\% & 67.80\% & 6.07\% & 64.68\% \\
		    2  &   $38720$ & 10.79\%& 82.80\% & 24.32\%&69.27\% & 1.01\% & 69.74\% & 5.82 \% & 64.93\% \\
		    3  &   $112576$ & 10.67\%& 82.92\%& 23.15\% &70.44\% & 1.38\% & 69.37\% & 5.82\% & 64.93\% \\
		    4  &   $127160$ &  9.73\%&  83.68\% & 22.55\% & 71.04\% &0.87\% & 69.88\% & 4.46\% & 66.29\% \\
     		 5  & $422328$    &  8.83\%& 84.76\%& 21.73\% & 71.86\% & 0.86\% & 69.89\% & 4.72\% & 66.03\% \\
     		 6  & $1012408$    & 9.64\%&  83.95\%& 20.36\% & 73.23\% & 1.00\% & 69.75\% & 3.50\% & 67.25\% \\
     		 7  & $1602488$    & 10.16\%& 83.43\%& 20.22\% & 73.37\% & 0.98\% & 69.77\% & 3.42\% & 67.33\% \\
     		 8  & $2782648$    & 9.53\%&  84.06\%& 20.00\% & 73.59\% & 1.00\% & 69.75\% & 2.93\% & 67.82\% \\
     		 9  & $5142456$    & 10.41\%&  83.18\%& 18.11\% & 74.89\% & 1.00\% & 69.75\% & 2.91\% & 67.84\% \\
     		 10  & $7502264$    & 10.05\%& 83.54\%& 18.13\% &74.92\% & 1.00\% & 69.75\% &2.52\% & 68.23\% \\ 
     		 11  & $9862072$    & 10.02\%&  83.57\%&17.48 \% &76.11 \% & 1.00\% & 69.75\% & 2.44\% & 68.31\% \\
     		 12  & $12221880$    & 9.93\%&  83.66\%& 17.29\% & 76.30\% & 1.00\% & 69.75\% & 2.42\% & 68.33\% \\
     		 13 & $14581688$    & 10.14\%&  83.45\%& 17.24\% &76.35 \% & 1.00\% & 69.75\% & 2.22\% & 68.53\% \\
     		 14  & $14844344$    & 10.03\%&  83.56\%&15.22 \% &78.37 \% & 1.00\% & 69.75\% & 2.12\% & 68.63\% \\
     		 15  & $14849474$    & 10.02\%&  83.57\%& 14.71\% & 78.88\% & 1.00\% & 69.75\% & 1.87\% & 68.88\% \\

     \hline
     %\bottomrule
\end{tabular} 
\end{table*}

In Fig. \ref{fig:permute}, the value of the diffusion parameter $n_{\text{d}}$ is fixed, and the model protection and anti-attack abilities are evaluated under various permute parameter settings, i.e. $n_{\text{p}}=\{1,2,3,4,5\}$. From Fig. \ref{fig:permute} (a) and (c), as the permute parameter $n_{\text{p}}$ increases, prediction accuracy drops more on both Mnist and Fashion-mnist datasets, which means higher model protection ability. Besides, encrypting more layers of CNN models leads to higher protection performance. Figure \ref{fig:permute} (b) and (d) present the ability to resist fine-tuning attacks on Mnist and Fashion-mnist datasets respectively, where the adversary has $1\%$ fraction of the original dataset. With the increasing permuting parameter, the CNN model has lower prediction accuracy after fine-tuning attacks, which means the encrypted models have better anti-attack ability and are better for model IP protection. Moreover, simple CNN has more encryption layers, and the encrypted model is more robust to fine-tuning attacks.

The impact of diffusion parameters $n_{\text{d}}$ on the performance of model protection and anti-attack ability is shown in Fig. \ref{fig:difussion}. The value of the permute parameter is fixed while the diffusion parameter is set in $\{1,2,3,4,5\}$. Figure \ref{fig:difussion} (a) and (c) present the prediction accuracy of the encrypted simple CNN model on Mnist and Fashion-mnist datasets under various diffusion parameters. Figure \ref{fig:difussion} (b) and (d) show the fine-tuning attack on the encrypted simple CNN model, which are evaluated on Minist and Fashion-mnist datasets, respectively. From the experimental results in Fig. \ref{fig:difussion}, the higher diffusion parameter $n_{\text{d}}$ leads to higher intelligent model protection and anti-attack abilities.

\subsubsection{Application on Complex Model}
To further evaluate the performance of the proposed intelligent model IP protection mechanism, we apply the proposed mechanism to the complex intelligent model VGG-16. We investigate and analyze the model protection and anti-attack abilities on both Cifar-10 and Cifar-100 datasets. The experimental results are presented in Tab. \ref{tab:vgg}. The original accuracy of the VGG-16 model on Cifar-10 and Cifar-100 are $93.59\%$ and $70.75\%$, respectively. As the encrypted layers and number of parameters increase, the prediction accuracy on Cifar-10 and Cifar-100 datasets converge to $10\%$ and $1\%$, respectively. Since the number of classifications in Cifar-10/Cifar-100 is $10$ and $100$, the encrypted VGG-16 model cannot get better results than a random guess, which are the minimum prediction accuracies from the information theory perspective. Besides, encrypting more layers and parameters of VGG-16 leads to stronger fine-tuning attack resistance and higher model robustness.

\section{Conclusion}
\label{sec:concl}

In this paper, we propose a PUF and permute-diffusion empowered AI model IP protection mechanism. First, we design the PUF and permute-diffusion encryption technologies based framework, where analyze the scenario and threat model of AI IP. Then, a PUF-based key generation protocol is designed. In this protocol, the delay-based Anderson PUF is adopted to generate the device-bind secret key. Besides, convolutional coding and interleaving technologies are combined to improve the stability of the PUF-based key generation and reconstruction. Next, a permute and diffusion-based weights encryption/decryption approach is proposed, where chaos theory is utilized to convert the PUF-based secret key to encryption/decryption keys. Moreover, the effectiveness of the proposed PUF and permute-diffusion empowered intelligent model IP protection mechanism is verified by the experimental evaluations. Future works are to propose the PUF-based intelligent model access control and user authentication methods.

% or


%\appendices
%% you can choose not to have a title for an appendix
%% if you want by leaving the argument blank
%\section{}

%\section*{Acknowledgment}
%
%
%This work is sponsored by the National Natural Science Foundation of China (Grant No.  61972255 and U20B2048).

\ifCLASSOPTIONcaptionsoff
  \newpage
\fi



\bibliographystyle{IEEEtran}
\bibliography{IEEEabrv,./bib/mylib}
%\begin{thebibliography}{1}
%\end{thebibliography}

\vspace{-9mm}
\begin{IEEEbiography}
[{\includegraphics[width=1in,height=1.25in,clip,keepaspectratio]{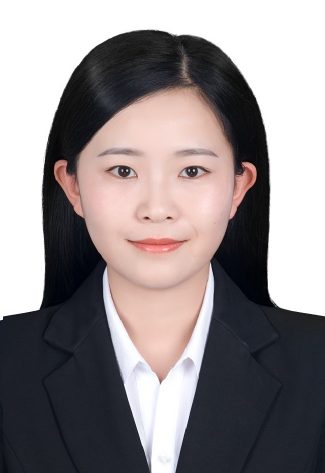}}]
{Qianqian Pan} received her B.S. and M.S. degree from the School of Information Science and Engineering, Southeast University, Nanjing, China, in 2015 and 2018, respectively. Now, she is pursuing the Ph.D. degree in the School of Electronic Information and Electrical Engineering, Shanghai Jiao Tong University, Shanghai, China. She is also a visiting Ph.D. student at Muroran Institute of Technology. Her research interests include security and privacy of machine learning, side-channel attacks, and so on. 
\end{IEEEbiography}

\vspace{-9mm}
\begin{IEEEbiography}
[{\includegraphics[width=1in,height=1.25in,clip,keepaspectratio]{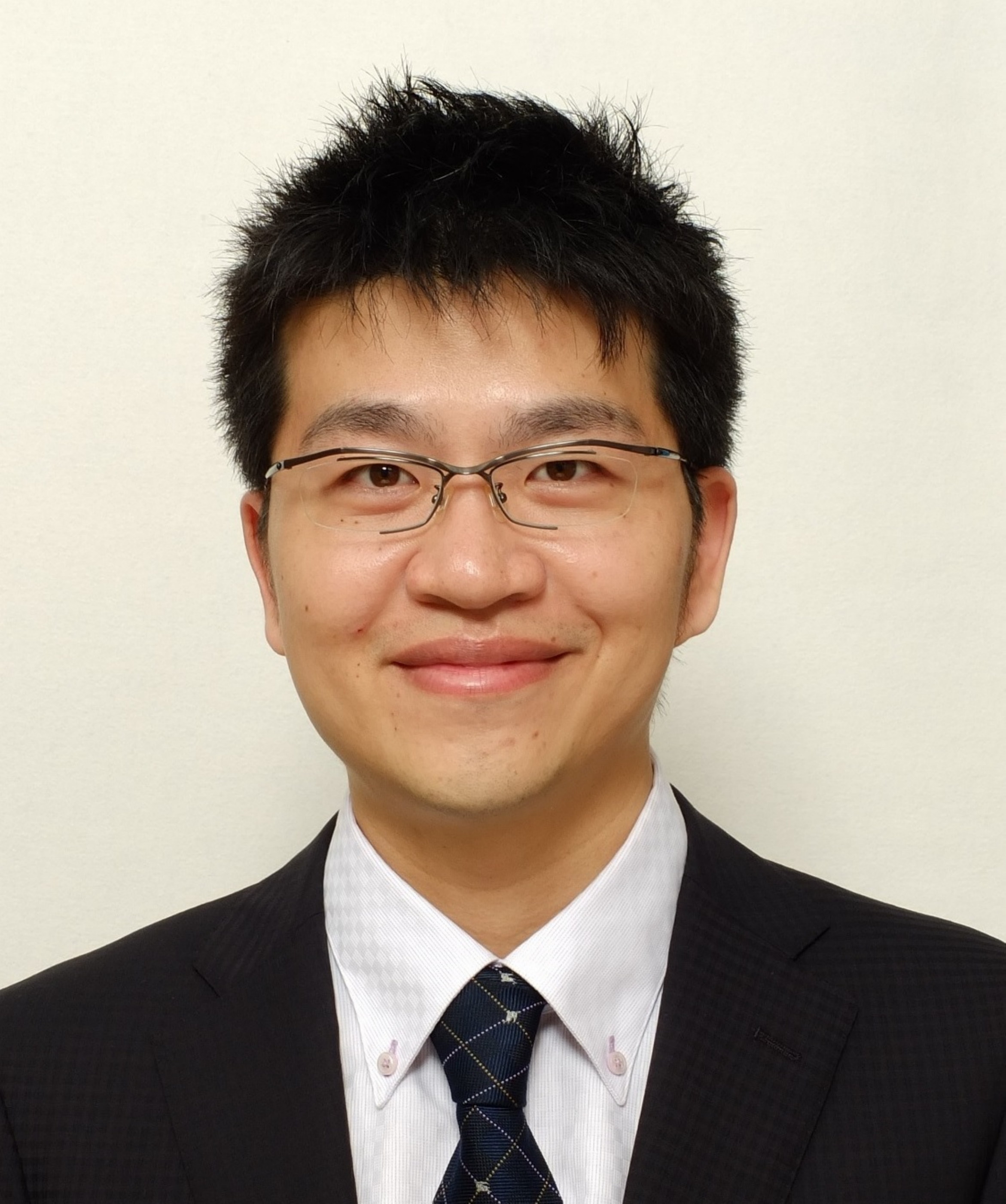}}]
{Mianxiong Dong} received B.S., M.S. and Ph.D. in Computer Science and Engineering from The University of Aizu, Japan. He is the Vice President and Professor of Muroran Institute of Technology, Japan. He was a JSPS Research Fellow with School of Computer Science and Engineering, The University of Aizu, Japan and was a visiting scholar with BBCR group at the University of Waterloo, Canada supported by JSPS Excellent Young Researcher Overseas Visit Program from April 2010 to August 2011. Dr. Dong was selected as a Foreigner Research Fellow (a total of 3 recipients all over Japan) by NEC C\&C Foundation in 2011. He is the recipient of The 12th IEEE ComSoc Asia-Pacific Young Researcher Award 2017, Funai Research Award 2018, NISTEP Researcher 2018 (one of only 11 people in Japan) in recognition of significant contributions in science and technology, The Young Scientists’ Award from MEXT in 2021, SUEMATSU-Yasuharu Award from IEICE in 2021, IEEE TCSC Middle Career Award in 2021. He is Clarivate Analytics 2019, 2021 Highly Cited Researcher (Web of Science) and Foreign Fellow of EAJ.
\end{IEEEbiography}

\begin{IEEEbiography}
[{\includegraphics[width=1in,height=1.25in,clip,keepaspectratio]{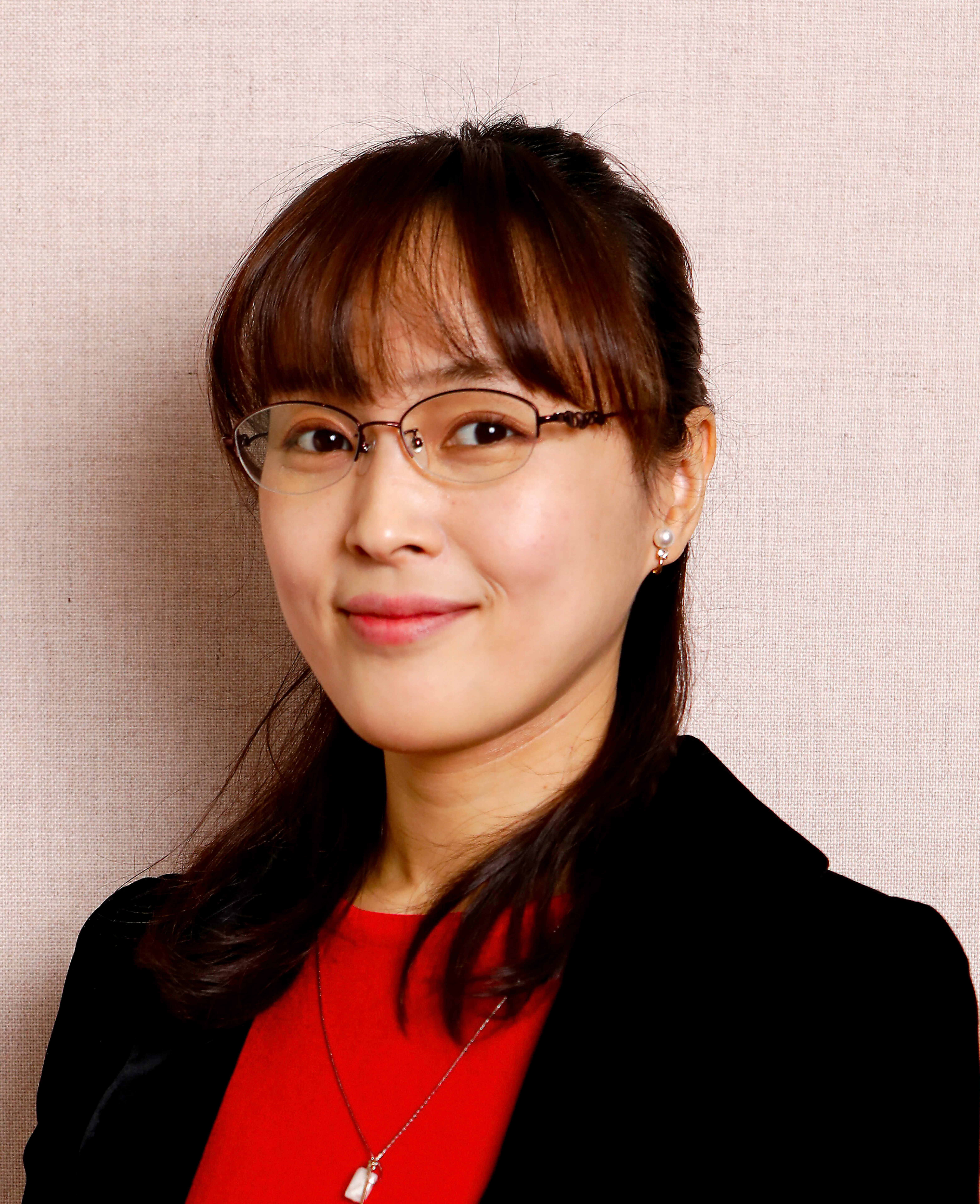}}]
{Kaoru Ota} was born in Aizu-Wakamatsu, Japan. She received M.S. degree in Computer Science from Oklahoma State University, the USA in 2008, B.S. and Ph.D. degrees in Computer Science and Engineering from The University of Aizu, Japan in 2006, 2012, respectively. Kaoru is a Professor and Ministry of Education, Culture, Sports, Science and Technology (MEXT) Excellent Young Researcher with the Department of Sciences and Informatics, Muroran Institute of Technology, Japan. From March 2010 to March 2011, she was a visiting scholar at the University of Waterloo, Canada. Also, she was a Japan Society of the Promotion of Science (JSPS) research fellow at Tohoku University, Japan from April 2012 to April 2013. Kaoru is the recipient of IEEE TCSC Early Career Award 2017, The 13th IEEE ComSoc Asia-Pacific Young Researcher Award 2018, 2020 N2Women: Rising Stars in Computer Networking and Communications, 2020 KDDI Foundation Encouragement Award, and 2021 IEEE Sapporo Young Professionals Best Researcher Award. She is Clarivate Analytics 2019, 2021 Highly Cited Researcher (Web of Science) and is selected as JST-PRESTO researcher in 2021, Fellow of EAJ in 2022.
\end{IEEEbiography}
\vspace{-7 mm}
\begin{IEEEbiography}
[{\includegraphics[width=1in,height=1.25in,clip,keepaspectratio]{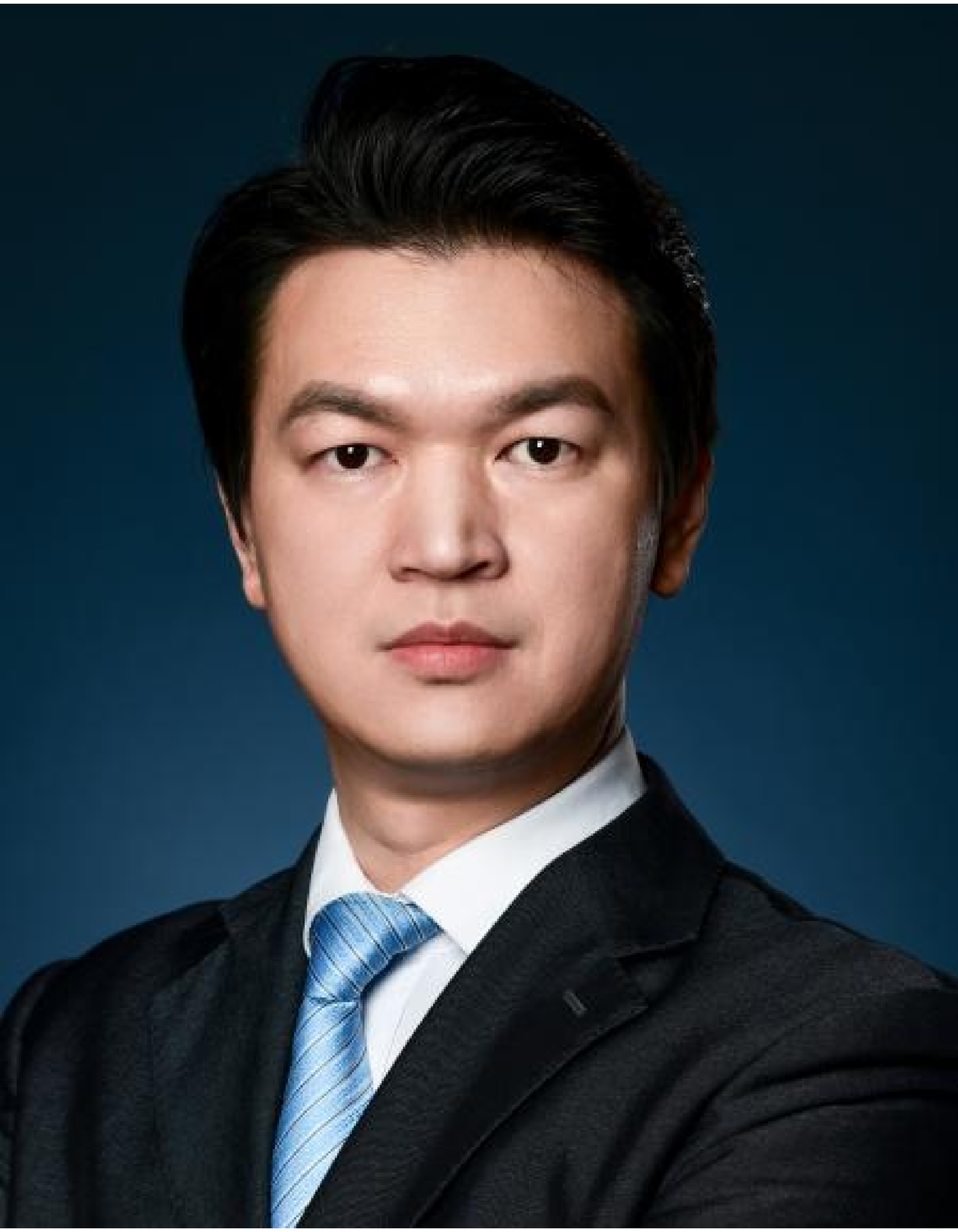}}]
{Jun Wu} received the Ph.D. degree in information and telecommunication studies from Waseda University, Japan, in 2011. He is currently a professor with the Graduate School of Information, Production and Systems of the same university. He is the chair of IEEE P21451-1-5 Standard Working Group for Internet of things. His research interests include the intelligence and security techniques of Internet of Things (IoT), edge computing, big data, 5G/6G, etc. He is the author or co-author of more than 200 peer-reviewed journal/conference papers within the above-mentioned topics. His publications have received a few distinctions, which includes the Best Paper Award of IEEE Transactions on Emerging Topics in Computing, in 2020, Best Paper Award of International Conference on Telecommunications and Signal Process in 2019, Best Conference Paper Award of the IEEE ComSoc Technical Committee on Communications Systems Integration and Modeling in 2018. He has served as the Track Chair for VTC 2019, VTC 2020 and the TPC Member of more than ten international conferences including ICC, GLOBECOM, etc. He severs an Associate Editor for the IEEE Systems Journal and IEEE Networking Letters. He has served the as a Guest Editor for the IEEE Transactions on Industrial Informatics, IEEE Transactions on Intelligent Transportation, IEEE Sensors Journal, Sensors, Frontiers of Information Technology \& Electronic Engineering (FITEE), etc.
\end{IEEEbiography}


%\vfill


\end{document}